\documentclass[12pt,a4paper]{article}
\usepackage{amssymb,amsmath,graphicx,subcaption,hyperref,cite}
\usepackage{epstopdf}
\usepackage[utf8]{inputenc}
\usepackage[english]{babel}

\begin{document}
	
	
\begin{center}
	{\Large\bf Metastability in graded and step like variation of field and anisotropy of the Blume-Capel ferromagnet }
\end{center}
\vskip 1 cm
\begin{center} %
	Moumita Naskar$^1$ and Muktish Acharyya$^2$
		
	\textit{Department of Physics, Presidency University,} 
		
	\textit{86/1 College Street, Kolkata-700073, India} 
	\vskip 0.2 cm
	\textit{Email$^1$:moumita1.rs@presiuniv.ac.in}
		
	\textit{Email$^2$:muktish.physics@presiuniv.ac.in}
\end{center}
\vspace {1.0 cm}
\vskip 0.5 cm
\noindent {\large\bf Abstract:}
The metastable behaviour of two dimensional anisotropic Blume-Capel ferromagnet, under the influence of graded and stepped like variations of magnetic field, has been investigated by extensive Monte Carlo simulation using Metropolis single spin flip algorithm. Starting from the initial perfectly ordered state, reversal of magnetisation has been studied in presence of the field. Metastable lifetime or the reversal time of magnetisation for the system having uniform anisotropy and in the presence of both graded and stepped field has been studied. Also the same has been explored for a {\it graded and stepped anisotropic system} in presence of a uniform field. Finite size effect is analyzed for the variation of reversal time with gradient of field ($G_h$) and a scaling relation $\langle \tau \rangle \sim L^{-\beta} f(G_h L^\alpha)$ is proposed. Spatial variation of density profile for the projection states (i.e., +1, 0 and -1) has also been studied at the moment of reversal of magnetisation. The spatial variation of the number of spin flips per site is studied. Motion of the interface (or domain wall) with the gradient of field and gradient of anisotropy are investigated and are found to follow the
hyperbolic tangent like behaviours in both cases. Since, both the anisotropy and the applied field have significant impact on the metastable lifetime, an interesting competitive scenario is observed for a {\it graded anisotropic system in graded field} and a {\it stepped anisotropic system in stepped field}. A line of marginal competition has been found out for both the cases separating the regions of field dominated reversal and anisotropy dominated reversal. The decay of the metastable volume fraction was found
to follow the Avrami's law. The  mean reversal time was observed to decrease monotonically across the phase boundary of the ferro-para transition.

\newpage
\vskip 0.5 cm
\noindent {\large\bf I. Introduction}
\vskip 0.5 cm

In the technology of magnetic recording\cite{techno,daniel}, the magnetisation switching time (or the reversal time) plays the crucial role, to determine the speed of recording. This prompted the theoretical and experimental researchers to study
the reversal phenomena with intense care. The reversal of magnetisation and the 
dependence of the reversal time on the applied magnetic field were historically
analysed by Becker-D\"{o}ring theory\cite{becker,gunton}. The theory was well tested
\cite{mastauff}
in large scale computer simulation in Ising ferromagnets using multispin coding techniques. To study the reversal mechanism and for better understanding of the microscopic spin flip, the Ising ferromagnet was chosen as the simplest model. The 
longevity of metastable states in kinetic Ising 
ferromagnet is studied\cite{rikvold1}
to find its dependence on the applied field and system size. The rates of growth and decay of the
clusters of different sizes, have been studied\cite{vehkamaki} as functions of external field and temperature. The
rate of nucleation of critical nuclei, its speed of expansion and the corresponding changes of free
energy is related and is described by famous 
Kolmogorov–Johnson–Mehl–Avrami law\cite{kolmogorov,johnson,avrami}. 

The switching in the environment of nonuniform parameters (field, temperature etc.) 
was investigated recently.
The heat assisted magnetisation reversal was investigated
\cite{rikvold2} in ultra thin films for ultra-high-density information
recording media. The statistical behaviours of magnetisation reversal was studied
\cite{naskar1}
in random field Ising model by Monte Carlo simulation to investigate the role of
quenched impurity in the switching phenomena. The reversal with nonuniform
(over space) magnetic
field was also investigated\cite{abyaya} and a marginally competitive line
was drawn\cite{ranajay} in the plane formed by the gradient of field and the gradient of temperature. The switching of magnetisation was investigated in Ising ferromagnets
by periodic pulsed \cite{bkc,ma2010} magnetic field and a time dependent field like
spreading Gaussian\cite{ma2014}. The relaxation of
Ising ferromagnet after a sudden reversal of applied magnetic field is 
also studied\cite{binder1}.

The above mentioned studies are basically done on Ising  model. In
the case of spin- 1/2 Ising model, the reversal is mainly governed by the 
microscopic spin flip.  Due to the application of external magnetic field, the spin
would flip its direction. The metastability and
its lifetime is function of the temperature and the magnitude of external magnetic field. {\it How does the reversal will mutate if the spin component ($S^z$) has a perpendicular component with respect to the direction of applied magnetic field
? }

Precisely,
how does the anisotropy take part in reversal mechanism in the spin-1 
anisotropic Blume-Capel 
\cite{blume,capel} model ?
Metastability and nucleation in the Spin-1 Blume-Capel (BC) ferromagnet was studied and found the different mechanism of transition\cite{cirillo}. 
The position of the tricritical point was determined by studying\cite{ipsita} the method of
geometric mutual information.The dynamical phase transition was
studied\cite{erol} in a randomly diluted single site anisotropic BC model in presence of time varying
oscillating magnetic field.
Recently, the role of single
site anisotropy to the reversal of Blume-Capel ferromagnet was 
investigated\cite{naskar2} by extensive Monte Carlo simulation.
The magnetic properties of ferrimagnetic mixed spin ($\sigma=2, S={7 \over 2}$) in a
Blume-Capel model was studied\cite{masrour} by Monte Carlo simulation. The behaviours of layered Blume-Capel model with
inhomogenous crystal field was also investigated\cite{bahmad} by extensive MC simulation and an interesting reentrant layering transition was found. The bilayered
graphyne structured BC model with RKKY interaction was studied\cite{fadil} by MC
simulation and found the increasing transition temperature by decreasing the number
of intermediate nonmagnetic layers. The zero field BC model was studied
\cite{nicos1} by Monte Carlo simulation (using a special hybrid algorithm) to study the behaviours (correlations) in the vicinity of
first order and second order phase boundary. The high precision MC techniques were
used\cite{nicos2} to study the universality class of random bond disordered BC model (in two dimensions) and concluded
that it belongs to the universality class of two dimensional Ising model (with additional logarithmic
corrections).

The thermally
activated magnetisation switching of small ferromagnetic particles driven by an external magnetic
field has been investigated and interestingly a crossover from coherent rotation to nucleation for a
classical anisotropic Heisenberg model, has been reported\cite{uli}.

Recently, in the field of experimental research in magnetisation switching, 
the nonuniform field and temperature were introduced. Temperature gradient-induced
magnetisation reversal of single
ferromagnetic nanowires was studied\cite{expt1} in $Co_{39}Ni_{61}$. The techniques of
confinement of magnetic nanoparticles (sub-100 nm) by using localised field gradients was reported\cite{expt2} recently. Continuously graded anisotropy in single 
Fe-Pt-Cu composites
films was incorporated experimentally to have a conclusion that an anisotropy
gradient can be realized, and tailored, in single continuous films without the need for multilayers\cite{expt3}.

These experimental results prompted us to study the role of continuous (graded) and discrete (step like) variation of anisotropy
in the magnetic model system of discrete spin symmetry (here BC model). What kind of 
the morphological changes are governing the reversal mechanism ? The ordering kinetics,
growth of clusters are well studied in the magnetic model system for uniform field and anisotropy. However, the role of spatial variation of field and anisotropy in the
reversal mechanism is not well known.
These are the main motivation of the present study. In this article, reversal is studied (by Monte Carlo simulation) in the Blume-Capel
ferromagnet having the graded field
, stepped field, graded anisotropy and stepped anisotropy. The manuscript is organised as follows: the model is introduced in the next section-II, the results and analysis
are reported in section -III and the paper ends with a summary in section-IV. 

\vskip 2 cm

\noindent {\large\bf II. The Model and Simulation method:}
\vskip 0.5 cm
The spin-1 Blume-Capel model can be described by the following Hamiltonian,
\begin{equation}
H = -J \sum_{<i,j>} \sigma_i^z \sigma_j^z + \sum_i D(i) (\sigma_i^z)^2 -  \sum_i h(i) \sigma_i^z
\end{equation}
where $\sigma_i^z$ is the component of spin along z-direction at i-th lattice site and 
can take values +1, 0 and -1. The first sum is restricted to the interaction between 
nearest neighbour spins only with uniform ferromagnetic interaction strength ($J>0$).
$J$ has the unit of energy. 
Second term considers the effect of single site anisotropy (or magneto-crystalline anisotropy) 
arising from crystal structure. $D(i)$ (measured in the unit of $J$) is the strength of the anisotropy experienced at 
i-th lattice site in the system. Third term tells us about the Zeeman energy involving 
the interaction of externally applied magnetic field $h(i)$ (applied in the z-direction and measured in the unit of $J$) with each individual 
spin. We have simulated a two dimensional ferromagnetic square lattice of size $L \times L$ 
with \textbf{open boundary condition} on the both directions. In this paper we have studied 
some behaviour of the system in presence of two different kinds of both the field and 
anisotropy.

$\bullet$ \textbf{\textit{Graded field and graded anisotropy:}} A gradient of field is 
applied here throughout the lattice sites (from left edge to right edge) along x-direction. 
So the form of the field is simply taken as,
\begin{equation}
h(x)= G_h \ast x + b
\end{equation}
where $ h(x) $ varies only along the x direction i.e. for a fixed $ x $, lattice sites along 
y direction are equifield sites. $G_h= \frac{dh}{dx}$ is the \textbf{gradient} of 
the field. So if a field of strength $h_l$ is applied at the left boundary of the 
lattice and $h_r$ at the right boundary, then the lattice experiences a gradient of 
field $G_h= \frac{h_r-h_l}{L}$ since $h(x)= b= h_l$ at $x= 0$ (left boundary) and 
$h(x)= h_r$ at $x= L$ (right boundary). In a similar way the gradient of anisotropy 
is also considered here,
\begin{equation}
D(x)= G_D \ast x + c
\end{equation}
where $ D(x) $ varies along the x direction i.e. for a fixed $ x $, lattice sites along y 
direction are equianisotropic sites. So the \textbf{gradient} of the anisotropy of the system 
is $G_D= \frac{D_r- D_l}{L}$ where $D_l$ is the anisotropy set at the left boundary of 
the lattice and $D_r$ is at right. $ D(x) $ is always taken positive throughout our work.    

$\bullet$ \textbf{\textit{Stepped field and stepped anisotropy:}} In that case, the field 
acts in a stepped way where half of the lattice sites along x direction are influenced 
by a field $h_{sl}$ (at left side) and another half are influenced by $h_{sr}$ (at right 
side).
\begin{equation}
h(x)= h_{sl} \;\;\;\; for \;\;\ 1 \le x \le \frac{L}{2}, \; \forall y
\end{equation}
$$ \;\;\;\;\;\; = h_{sr} \;\;\;\; for \;\; \frac{L}{2} < x \le L, \; \forall y$$
 Similarly as graded field, for a fixed $ x $ all the $ h(x) $ along y direction are same. \textbf{Step 
 difference} ($ S_h $) is simply calculated by, $S_h= |h_{sr} - h_{sl}|$. In the same way, 
 stepped anisotropy is,
 \begin{equation}
 D(x)= D_{sl} \;\;\;\; for \;\;\ 1 \le x \le \frac{L}{2}, \; \forall y
 \end{equation}
 $$ \;\;\;\;\;\;\; = D_{sr} \;\;\;\; for \;\; \frac{L}{2} < x \le L, \; \forall y$$ 
  if half of the lattice sites at left side have the strength of anisotropy $D_{sl}$ and 
  half at the right side have $D_{sr}$ then \textbf{step difference} is simply $S_D= |D_{sr} - D_{sl}|$. 

Initially the system is considered to be in perfectly ordered state ($ T<T_c $) where 
all the spins are $\sigma_i^z = +1 \;\;\forall i$. Lets discuss the way lattice is 
updated (by random updating scheme). A lattice site (i-th say) has been chosen randomly. 
Say initially the spin at that site is $\sigma_i^z(initial)$. The updated spin state may 
be any of the three states (+1, 0 and -1) which has been determined with equal probability 
in random way. Let the final state be $\sigma_i^z(final)$. Now whether the spin 
$\sigma_i^z(initial)$ will be updated to the state $\sigma_i^z(final)$ ultimately that is 
decided by the Metropolis transition probability \cite{metrop},
\begin{equation}
P\big(\sigma_i^z(initial) \rightarrow \sigma_i^z(final)\big)= Min\big[1,e^{-\frac{\Delta H}{k_B T}}\big]
\end{equation}
where $\Delta H$ is the change in energy (measured in the unit of $J$) due to the change in spin state from 
$\sigma_i^z(initial)$ to $\sigma_i^z(final)$.  $k_B$ is the Boltzmann constant and $ T $ is the 
temperature (measured in the unit of $J/{k_B}$) of the system. According to the Metropolis algorithm\cite{metrop} the probability of spin flip
 $ P\big(\sigma_i^z(initial) \rightarrow \sigma_i^z(final)\big) $ 
is determined by calculating $\Delta H$ from the Hamiltonian and then compared it to a 
random number $r_n$ (uniformly distributed in the range [0:1]). The final state 
$\sigma_i^z(final)$ is accepted if $P \ge r_n$ otherwise the system is considered to be 
in initial state $\sigma_i^z(initial)$. In this way total $L^2$ number of 
randomly chosen lattice sites are updated and thus one Monte Carlo Step per Spin (MCSS) is 
completed which acts as the unit of time throughout the whole study. At each MCSS, the 
magnetisation of the system is determined by,
\begin{equation}
m(t)= \frac{1}{L^2} \sum_{i}^{L^2} \sigma_i^z
\end{equation}
$ t $ is the time in unit of MCSS.
\vskip 2 cm
\noindent {\large\bf III. Simulational Results} 
\vskip 0.5 cm

\begin{figure}[h!]
\begin{center}
\begin{subfigure}{0.5\textwidth}
	\includegraphics[angle=-90,width=\textwidth]{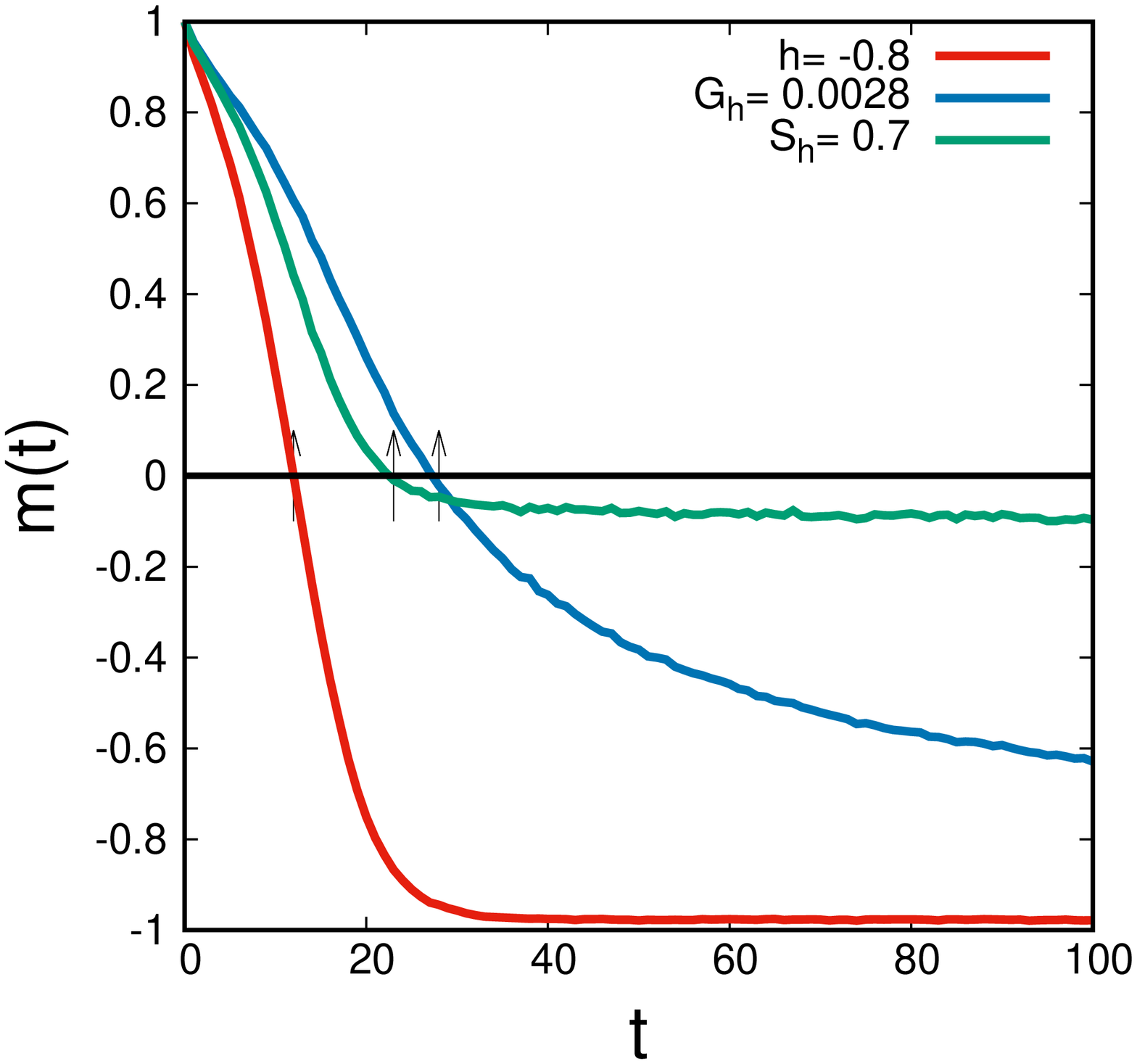}
	\subcaption{}
\end{subfigure}
\end{center}
\begin{subfigure}{0.5\textwidth}
		\includegraphics[angle=-90,width=0.9\textwidth]{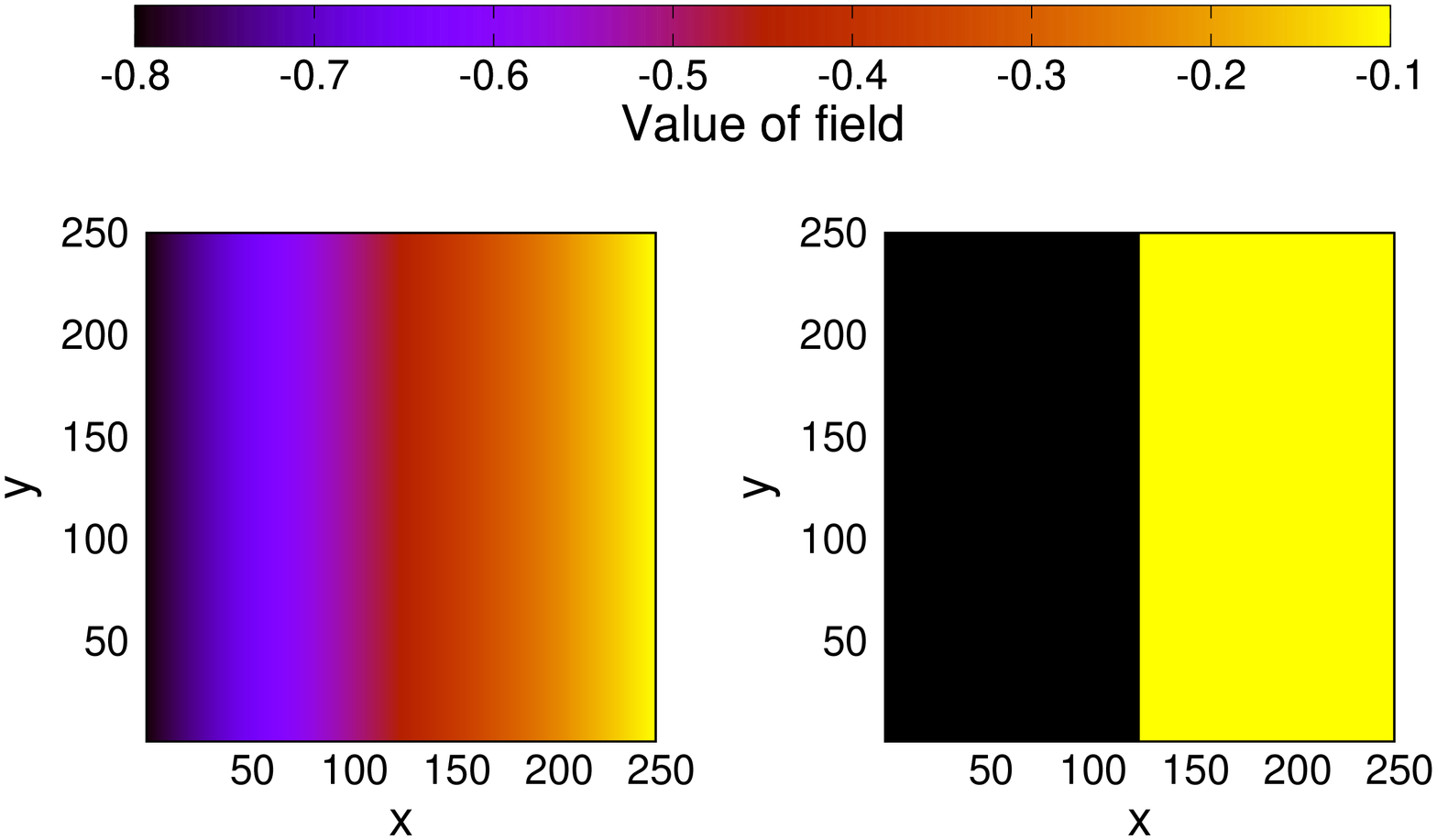}
		\subcaption{}
	\end{subfigure}
\begin{subfigure}{0.5\textwidth}
	\includegraphics[angle=-90,width=0.9\textwidth]{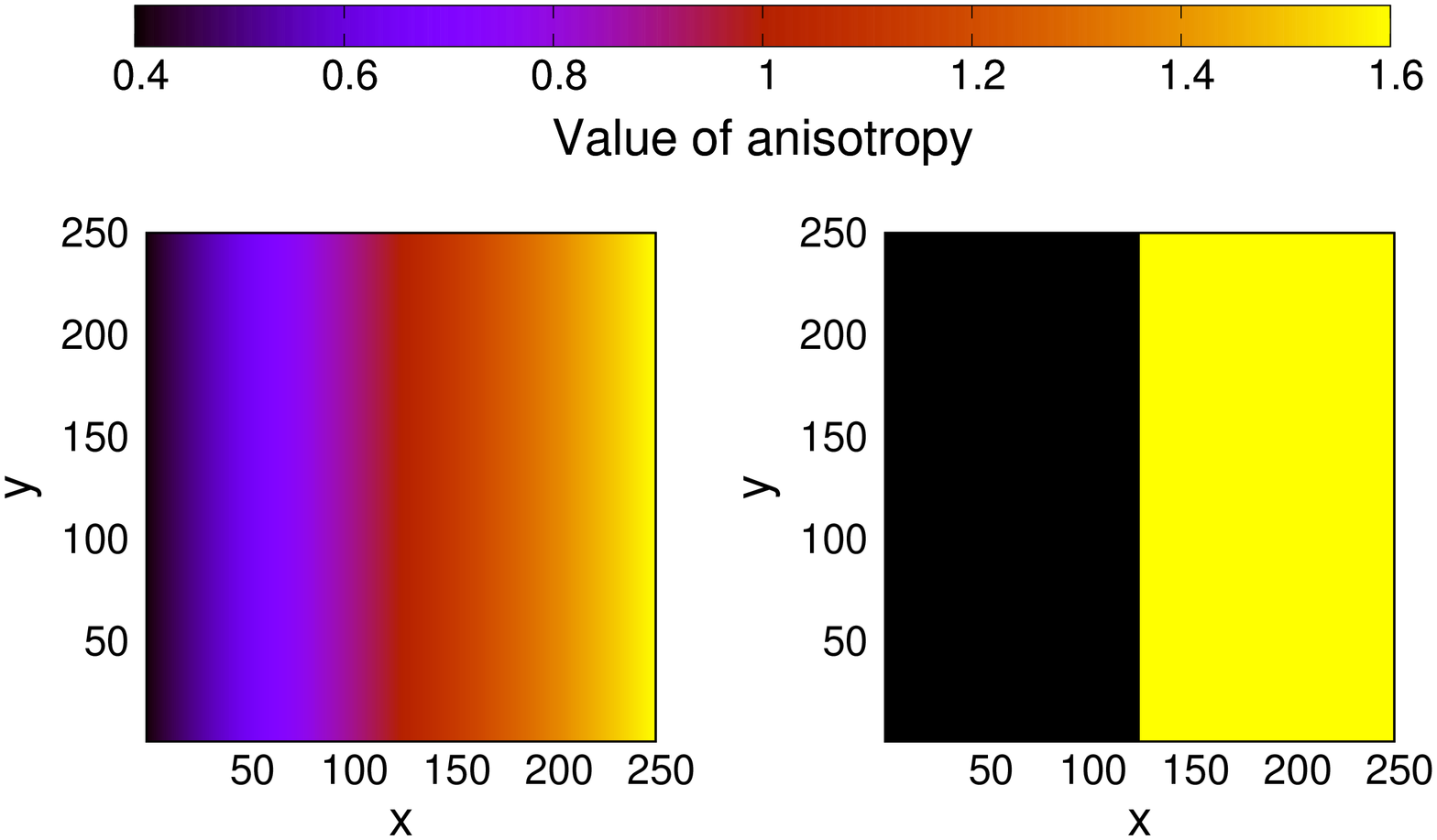}
	\subcaption{}
\end{subfigure}

	\caption{(a) Variation in magnetisation (m(t)) with time (t) in presence of a uniform 
		field $h=-0.8$ (red line, $\tau=12$ MCSS), gradient of field (blue line; 
		$h_l=-0.8$, $h_r=-0.1$, $G_h=0.0028$, $\tau=28$ MCSS) and stepped field  
		(green line; $h_{sl}=-0.8$, $h_{sr}=-0.1$, $S_h= 0.7$, 
		$\tau=23$ MCSS). 		
		(b) Image plot of the applied \textbf{Left:} graded field having 
		gradient $G_h= 0.0028$ ($h_l= -0.8$, $h_r= -0.1$). \textbf{Right:} stepped field of step 
		difference $S_h= 0.7$ ($h_{sl}= -0.8$, $h_{sr}= -0.1$). Anisotropy of the system is $\textbf{D= 1.6}$. Temperature is 
		kept fixed at $\textbf{T= 0.8}$.(c)  Image plot of the \textbf{Left:} graded anisotropic system having 
		gradient $G_D= 0.0048$ ($D_l= 0.4$, $D_r= 1.6$). \textbf{Right:} stepped anisotropic system of step 
		difference $S_D= 1.2$ ($D_{sl}= 0.4$, $D_{sr}= 1.6$). }
\label{magtime}
\end{figure}

At $T<T_c$ starting from the perfectly ordered state ($m(t)=1$, all the spins are +1), 
variation of magnetisation with time is studied (fig-\ref{magtime}a) under the influence of 
three different kinds of external field applied in the opposite direction. An anisotropic system having a strength of  
anisotropy $D=1.6$ is considered here. In the presence of a uniform magnetic field, system remains in metastable state up to a certain period of time, beyond which large number of spins flip towards the  direction of applied field and eventually reaches a true equilibrium state ($m(t)$ close to -1).  Lets define the time 
taken by the system by which magnetisation starts to acquire negative value ($m(t)< 0$
but $m(t) \simeq 0$) 
as the \textbf{reversal time or metastable lifetime} $\tau$ of magnetisation of the system. Black arrows denote the 
reversal time for each case. Red curve presents the evolution of $ m(t) $ in presence of a 
uniform field of strength $h= -0.8$ interacting with the spin at each lattice site. Now a 
small gradient of field $G_h= 0.0028$ is generated along the right edge of the lattice by 
applying the field $h_l= -0.8$ at the left boundary and $h_r= -0.1$ at the right boundary of 
the lattice. Since we are dealing with negative field,  in presence of such $G_h$, the field
strength (or magnitude of the field) actually decreases from $h_l= -0.8$ gradually along the right side (positive x-direction). 
So obviously the reversal time rises (blue line) in that case in comparison to the case for 
uniform field. In presence of stepped field (green curve), $h_{sl}= -0.8$ (taken same as $h_l$ as considered in the case of graded field) 
is applied to the half of the lattice sites (left side) and $h_{sr}= -0.1$ (taken same as $h_r$ as considered in the case of graded field) 
is applied to the right half. For the presence of $h_{sr}= -0.1$ at the right half lattice sites, 
definitely the reversal time will be higher than that for uniform field. In addition, we have 
also noticed that $\tau$ for stepped field is a little bit smaller compared to that for graded 
field.

In fig-\ref{magtime}b image-plot of the two types of applied magnetic field used in the 
fig-\ref{magtime}a has been illustrated. Left imageplot of fig-\ref{magtime}b reflects the presence of a graded field 
having gradient $G_h= 0.0028$ ($h_l=-0.8$; $h_r=-0.1$) along the right side. Right imageplot of fig-\ref{magtime}b 
reflects the presence of a step of the applied field. Left half lattice sites are excited by 
$h_{sl}= -0.8$ (black colour) and the right half sites are excited by $h_{sr}= -0.1$ 
(yellow colour). Similarly fig-\ref{magtime}c represents the distribution of anisotropy across the lattice. Left one illustrates the graded anisotropic system with $D_l= 0.4$ and $D_r= 1.6$ and the right one presents steplike anisotropic system with $D_{sl}= 0.4$ and $D_{sr}= 1.6$. However, in the study of position of interface for a graded anisotropic system opposite direction of $G_D$ is considered that means $D_l= 1.6$ and $D_r$ is varied from $1.0$ to $0.4$.  

\begin{figure}[h!]
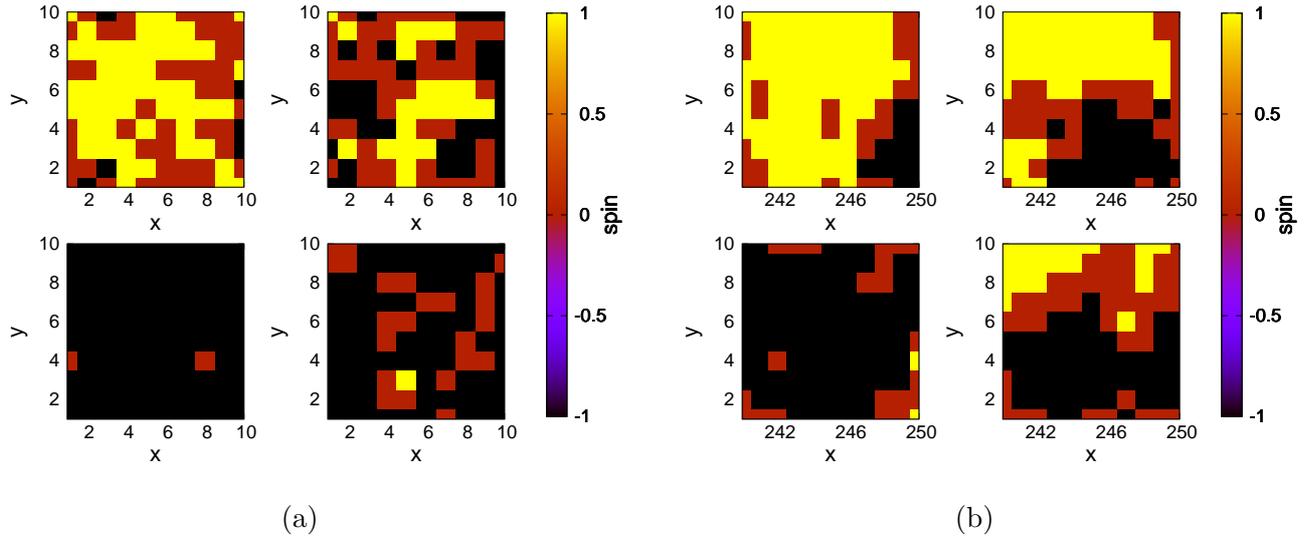

	\begin{subfigure}{0.5\textwidth}
        \includegraphics[angle=-90,width=\textwidth]{snap_lc.eps}
		\subcaption{}
	\end{subfigure}
	\begin{subfigure}{0.5\textwidth}
        \includegraphics[angle=-90,width=\textwidth]{snap_rc.eps}
		\subcaption{}
	\end{subfigure}	
	\caption{(a) In presence of gradient of field $G_h= 0.0028$, snapshots of the \textbf{top left} corner ($10\times10$ microlattice)
		 of the main lattice are 
		 taken at several time steps (arranged \textbf{clockwise} starting from top left). \textit{Top left:} $\tau= 9$, \textit{top right:} $\tau= 14$, \textit{bottom right:} 
		 $\tau= 24$, \textit{bottom left:} $\tau=32$ MCSS. Clearly the reversal occurs via the coalescence of many droplets of spin '-1'. (b) Under $G_h=0.0028$, snapshots of the \textbf{top right} corner ($10\times10$ microlattice) 
		 of the main lattice are taken 
		 at several time steps (arranged \textbf{clockwise} starting from top left). \textit{Top left:} $\tau= 360$, \textit{top right:} $\tau= 380$, \textit{bottom right:} 
		 $\tau= 423$, \textit{bottom left:} $\tau=500$ MCSS. Here the single nucleating cluster grows. Anisotropy of the system is $\textbf{D= 1.6}$. 
		 Temperature is $\textbf{T= 0.8}$.}
	\label{micro}
\end{figure}

Lets focus on the spin reversal mechanism responsible for the reversal of magnetisation in 
presence of a graded field. A gradient of field $G_h= 0.0028$ ($h_l=-0.8$; $h_r=-0.1$) acting 
towards right boundary is applied to the system. Now if we consider a micro-lattice of size 
$10 \times 10$ at the top left (named 'A') and top right corner (named 'B') of the main 
lattice then the field at the left edge of the microlattice A is $(h_l)_A= -0.8$ and that at 
right edge is $(h_r)_A= -0.772$. Same for the microlattice B will be $(h_l)_B= -0.128$ and 
$(h_r)_B= -0.1$. So the effective field acting on such microlattices can be assumed as almost 
uniform field. In microlattice-A (fig-\ref{micro}a), due to the strong field, reversal occurs 
through the formation of many clusters of spin '-1' (multi-droplet regime according to the 
Becker-Doring analysis of classical nucleation \cite{gunton}) results in relatively \textbf{shorter $\tau$}. 
Comparatively for microlattice-B, (fig-\ref{micro}b) reversal takes place by the growth of a single droplet (nucleation or single droplet regime) of spin '-1' causing a relatively 
\textbf{higher $\tau$}. So when applying such a graded field to the lattice, spins have greater 
tendency to get flipped on the left side first.

\begin{figure}[h!]
	\begin{subfigure}{0.5\textwidth}
		\includegraphics[angle=-90,width=\textwidth]{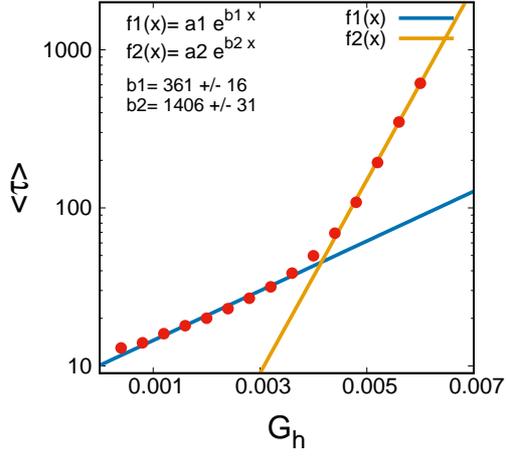}
		\subcaption{}
	\end{subfigure}
	\begin{subfigure}{0.5\textwidth}
	\includegraphics[angle=-90,width=\textwidth]{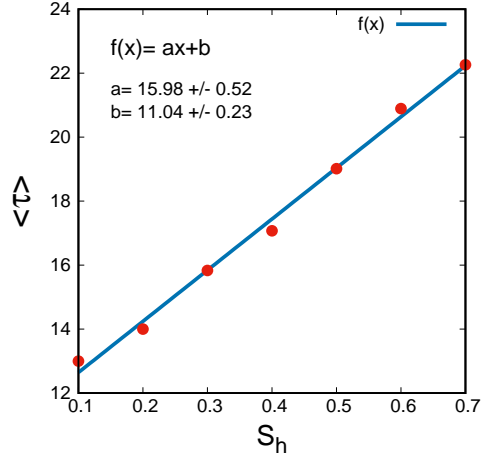}
	\subcaption{}
\end{subfigure}
\caption{(a) Variation of mean reversal time $\langle \tau \rangle$ with the strength of gradient 
	of field $G_h$ in semi-logarithmic scale.  To vary $G_h$, $h_l$ is kept fixed at $h_l= -0.8$ and $h_r$ is 
	varied from $h_r= -0.7$ to $+0.7$.  (b) Variation of reversal time $\langle \tau \rangle$
	with the step difference $ S_h $ of the  stepped field. To vary $S_h$, $h_{sl}$ is kept fixed at $h_{sl}= -0.8$ and $h_{sr}$ is varied from 
	$h_{sr}= -0.7$ to $-0.1$. Anisotropy of the system is $\textbf{D= 1.6}$. 
	Temperature is $\textbf{T= 0.8}$.}
 \label{revtime_h}
\end{figure}

\begin{figure}[h!]
	\begin{subfigure}{0.5\textwidth}
		\includegraphics[angle=-90,width=\textwidth]{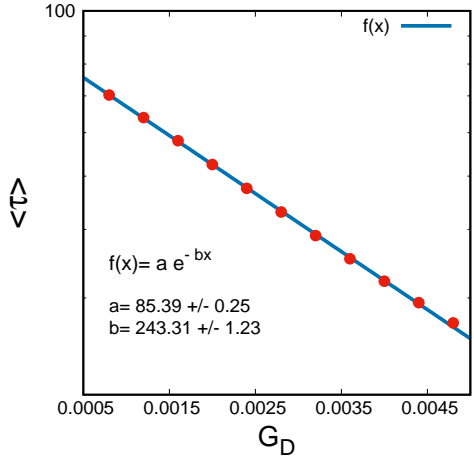}
		\subcaption{}
	\end{subfigure}
	\begin{subfigure}{0.5\textwidth}
		\includegraphics[angle=-90,width=\textwidth]{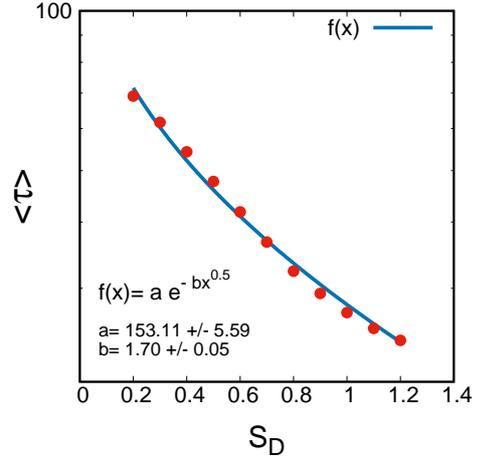}
		\subcaption{}
	\end{subfigure}
	\caption{(a) Semi-logarithmic plot of the variation of the mean reversal time $\langle \tau \rangle$ with the gradient 
		of anisotropy $G_D$. To vary $G_D$, $D_l$ is kept fixed at $D_l= 0.4$ and $D_r$ is 
		varied from $D_r= 0.6$ to $1.6$. (b) Variation of $\langle \tau \rangle$ 
		with the step difference of stepped anisotropy $S_D$. $ D_{sl} $ is kept fixed at $D_{sl}= 0.4$ and $D_{sr}$ is varied from 
		$D_{sr}= 0.6\;to\;1.6$ to vary $S_D$. Applied uniform field is kept fixed at $\textbf{h= - 0.8}$. Temperature is $\textbf{T= 0.8}$.}
	\label{revtime_D}
\end{figure}

We have studied the variation of average reversal time $\langle \tau \rangle$, obtained from randomly different  
1000 samples, with the strength of the gradient of field $G_h$ directed towards right edge. At 
the left boundary applied field is kept fixed at $h_l= -0.8$ while that at right boundary is 
varied from $h_r= -0.7$ to $+0.7$ to vary the $G_h$. $\langle \tau \rangle$ is found to 
increase exponentially (fig-\ref{revtime_h}a) with the increase in $G_h$ and also the presence of 
a crossover (in the rate of exponential growth) is noticed. Actually whenever the $h_r$ is set 
positive to produce a stronger 
$G_h$, some of the lattice sites near right edge do not participate in the reversal resulting 
a higher reversal time. That seems to be the reason of the appearance of the crossover. 
$\langle \tau \rangle$ is also studied (fig-\ref{revtime_h}b) with the step difference $S_h$ of 
the applied stepped field and it follows a straight line $\langle \tau \rangle \sim 15.98 S_h$. 
Step difference is varied here by keeping the field at the left half sites fixed at 
$h_{sl}= -0.8$ and varying at right half from $h_{sr}= -0.7\;\; to\; -0.1$. The linear variation 
of $\langle \tau \rangle$ is probably the reflection of the regular modulation of $h_{sr}$. 
Same study has been explored over the graded and stepped anisotropic system under the influence 
of a uniform field. In presence of a gradient of anisotropy (fig-\ref{revtime_D}a) acting towards right 
edge, reversal time decreases exponentially ($ \langle \tau \rangle \sim e^{- 243\; G_D} $) with 
the increase in $G_D$. Though the constant factor (243) is quite large in the exponential, we have to keep in mind the value of $G_D$, which is very small. In stepped anisotropic system (fig-\ref{revtime_D}b), variation of 
$\langle \tau \rangle$ with $S_D$ follows a stretched exponential $\langle \tau \rangle \sim e^{-1.7 (S_D)^{0.5}}$.

\begin{figure}[h!]
	\begin{subfigure}{0.5\textwidth}
	\includegraphics[angle=-90,width=\textwidth]{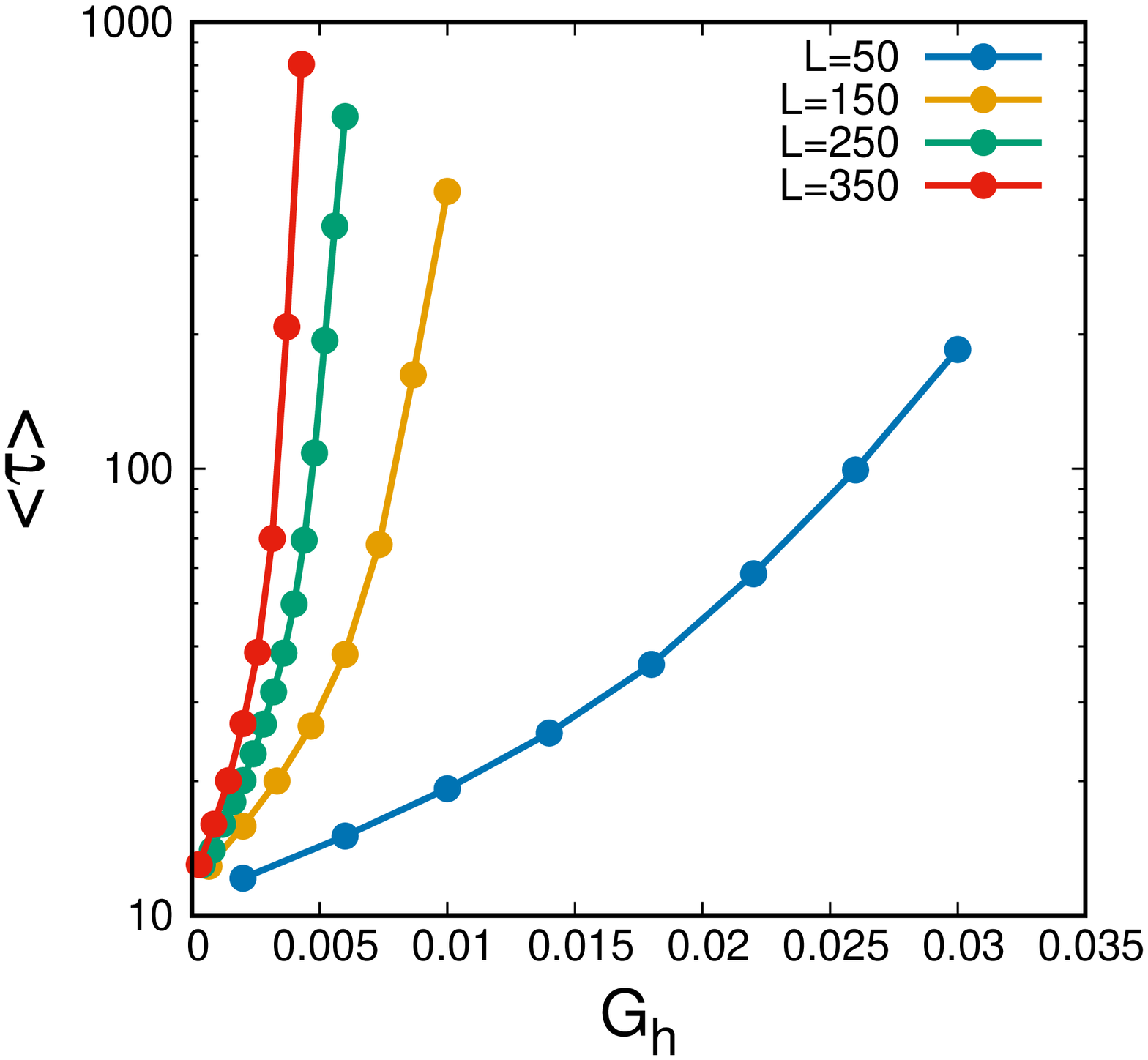}
	\subcaption{}
\end{subfigure}
\begin{subfigure}{0.5\textwidth}
	\includegraphics[angle=-90,width=\textwidth]{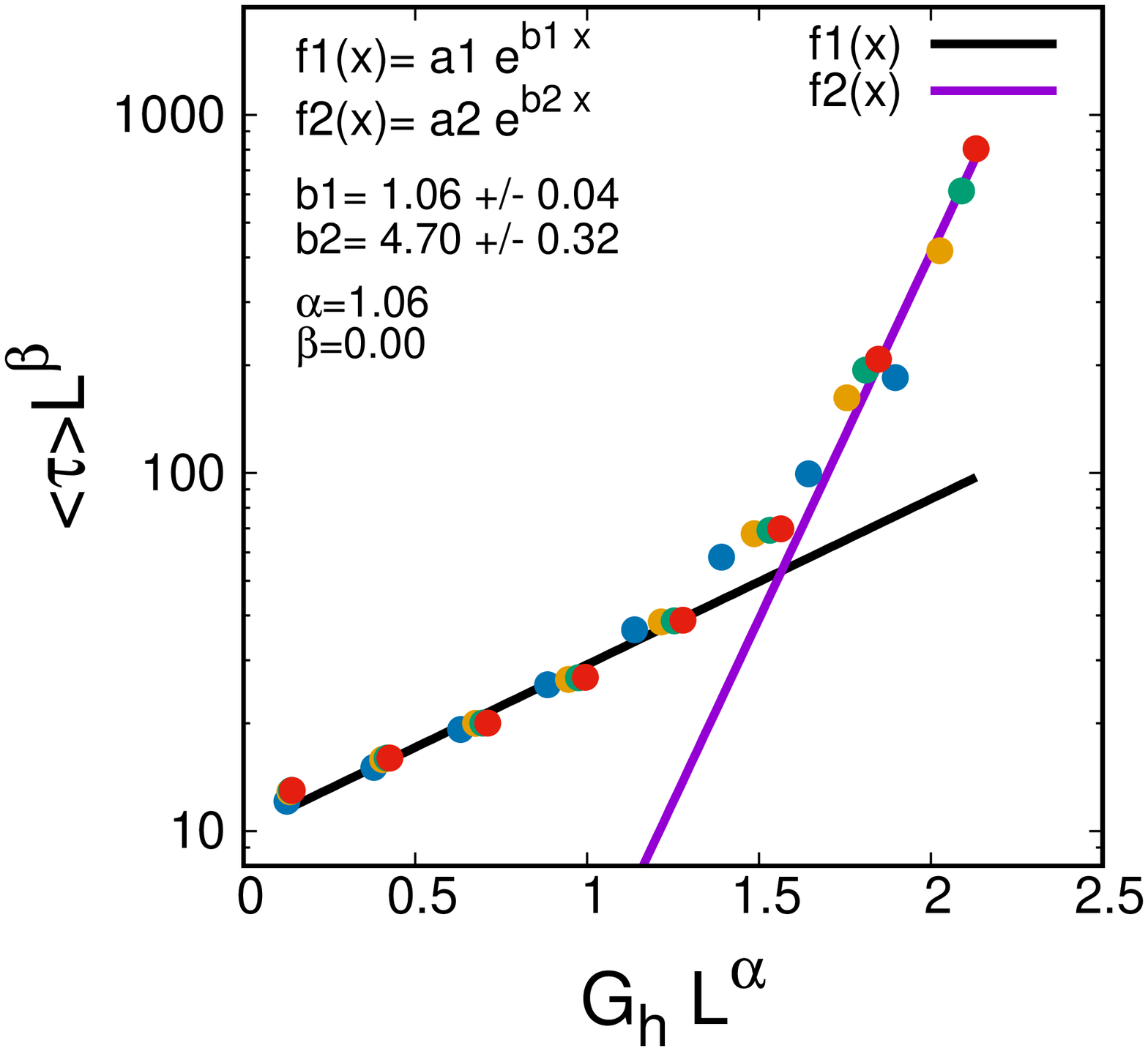}
	\subcaption{}
\end{subfigure}
	\caption{(a) Variation of $\langle \tau \rangle$ with $G_h$ for 4 different lattice 
		sizes ($L= 50, 150, 250\; and \;350$). Data are simply joined. Anisotropy of the system is fixed at 
		$\textbf{D= 1.6}$. Temperature is $\textbf{T= 0.8}$. (b) Data collapse of the first 
		plot by the exponent $\boldsymbol{\alpha= 1.06}$ and $\boldsymbol{\beta= 0}$ which has been fitted to the exponential function separately in 2 regions.}
	\label{fsize}
\end{figure}

\begin{figure}[h!]
	\centering
	\includegraphics[angle=-90, width= 0.6\textwidth]{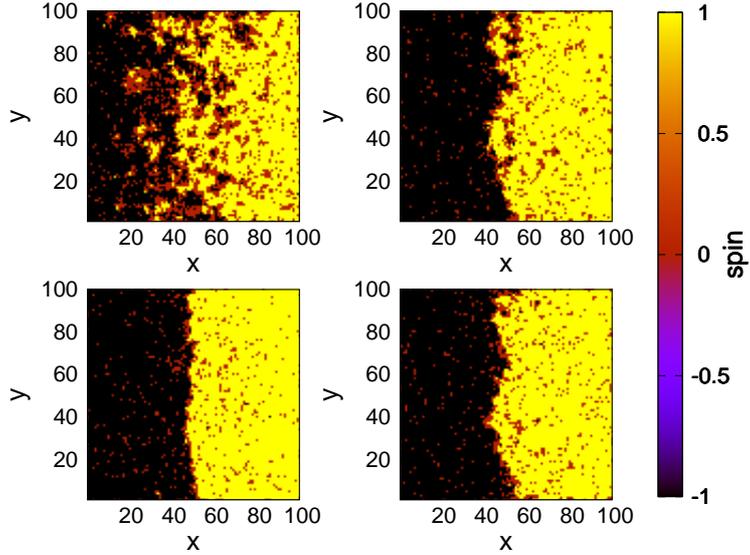}
	\caption{Imageplot (arranged \textbf{clockwise} starting from top left) of the lattice of size $L=100$ at the time of reversal for 4 different values of scaled gradient of field $G_h L^\alpha$ where $\alpha=1.06$ from previous scaled plot. \textbf{Top left:} $G_h L^\alpha = 1.0$; $\tau= 28$ MCSS, \textbf{Top right:} $G_h L^\alpha = 1.5$; $\tau= 73$ MCSS, \textbf{Bottom right:} $G_h L^\alpha = 1.6$; $\tau= 95$ MCSS, \textbf{Bottom left:} $G_h L^\alpha = 2.0$; $\tau= 313$ MCSS.}
	\label{snap_fsize}
\end{figure}

Since we are dealing with a lattice of small finite size ($L=250$), the system must suffer 
from the finite size effect. For that, variation of $\langle \tau \rangle$ with $G_h$ has been 
investigated (fig-\ref{fsize}a) for 4 different system sizes (data in the plot are simply joined). 
The system exhibits same kind of behaviour for each case. Additionally $\langle \tau \rangle$ 
shows a scaling relation $\langle \tau \rangle \sim L^{-\beta} f(G_h L^\alpha)$ where all the data for 
different lattice sizes are collapsed\cite{collapsed} (fig-\ref{fsize}b) to a single plot by the exponents 
$\alpha \simeq 1.06$ and $\beta = 0$. We have used our code to get the data collapse
by several trials of the values of the exponents to have visibly good collapse. The scaled plot is now fitted to the exponential function separately in two regions. Some snapshots (fig-\ref{snap_fsize}) of a single sample (of size $L=100$) at $\tau$ are taken for 4 different values of $G_h L^\alpha$ from which it can be inferred that the crossover appears when the interface becomes smoother.   

\begin{figure}[h!]
	\centering
	\begin{subfigure}{0.4\textwidth}
		\includegraphics[angle=-90,width=\textwidth]{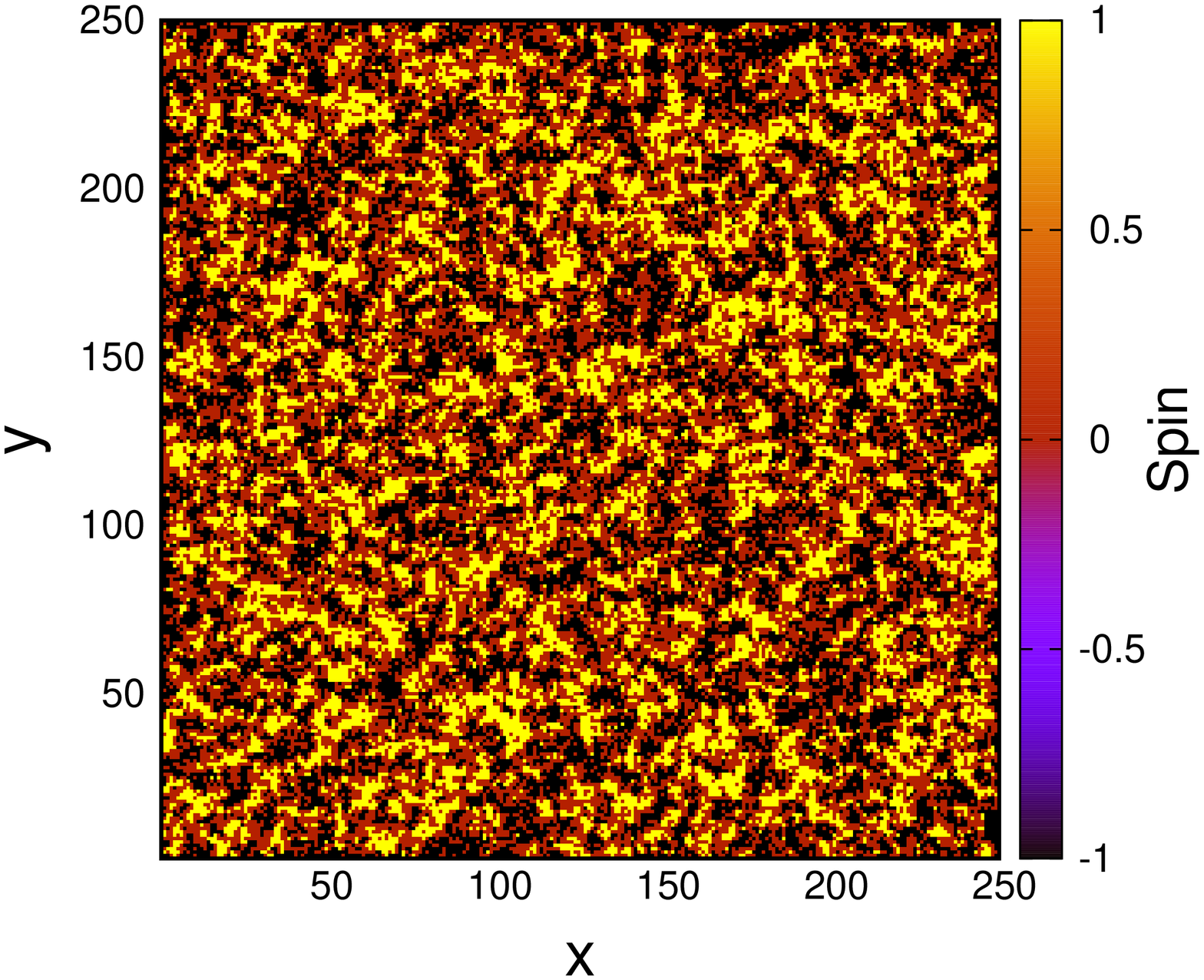}
		\subcaption{}
	\end{subfigure}
	\begin{subfigure}{0.4\textwidth}
		\includegraphics[angle=-90,width=\textwidth]{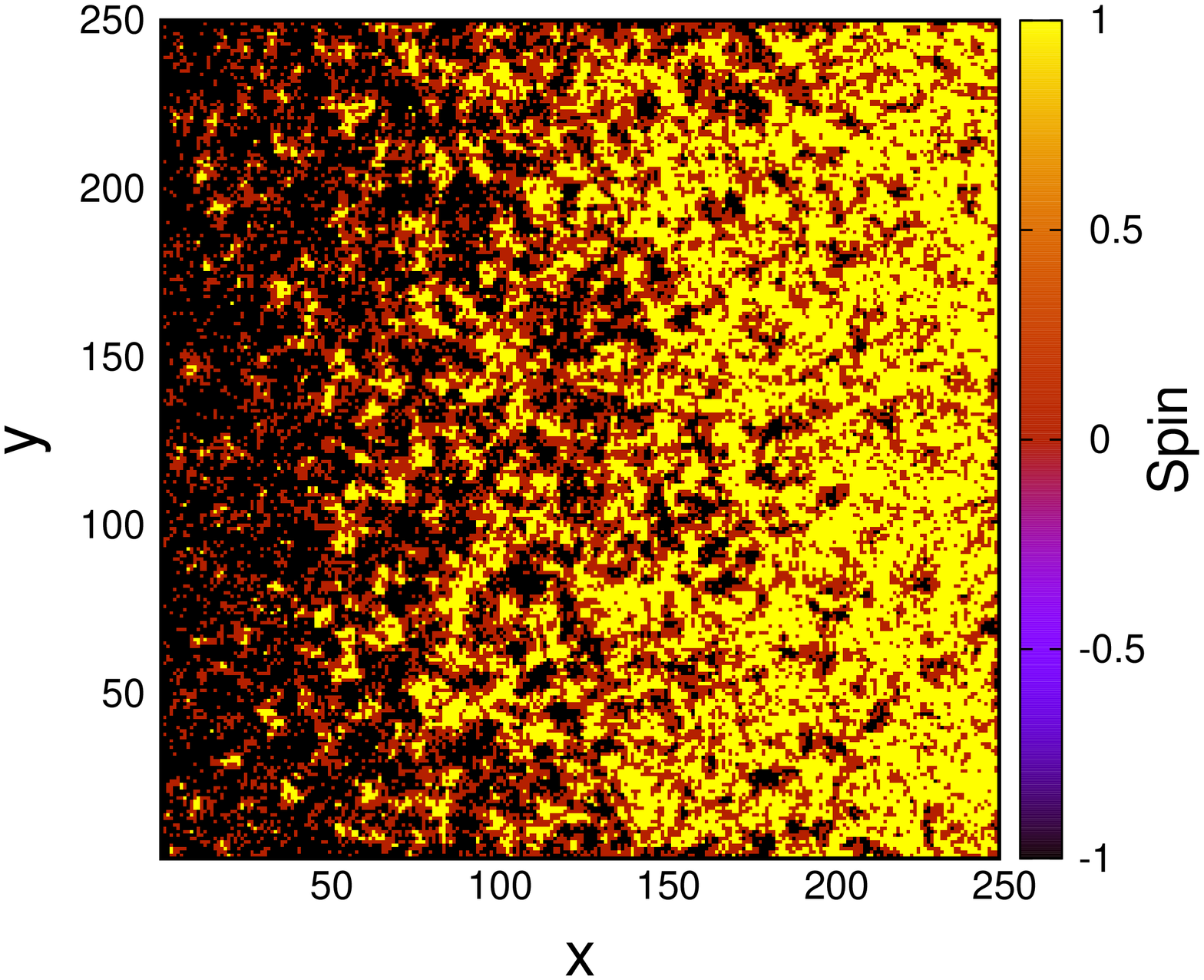}
		\subcaption{}
	\end{subfigure}	
	\begin{subfigure}{0.4\textwidth}
		\includegraphics[angle=-90,width=\textwidth]{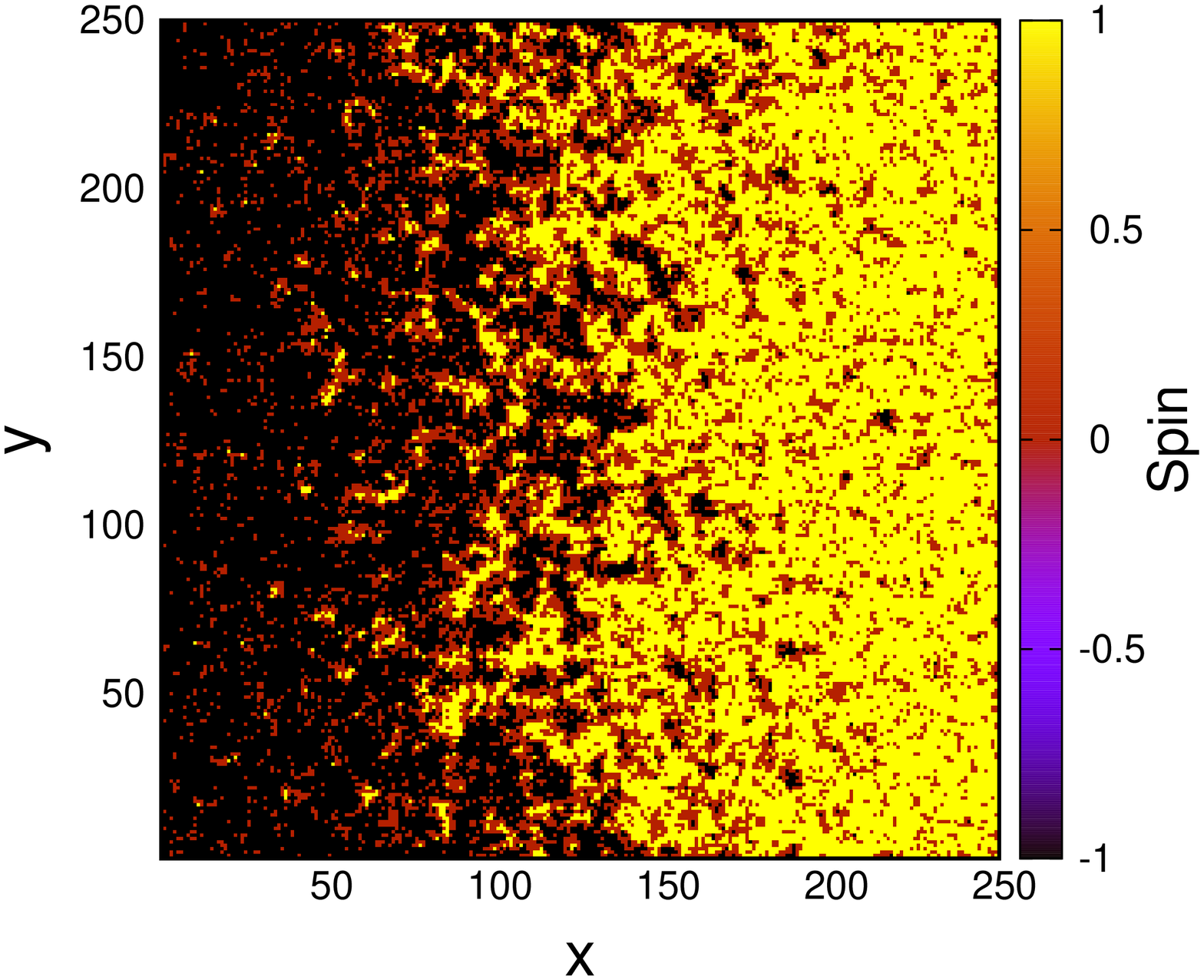}
		\subcaption{}
	\end{subfigure}
	\begin{subfigure}{0.4\textwidth}
		\includegraphics[angle=-90,width=\textwidth]{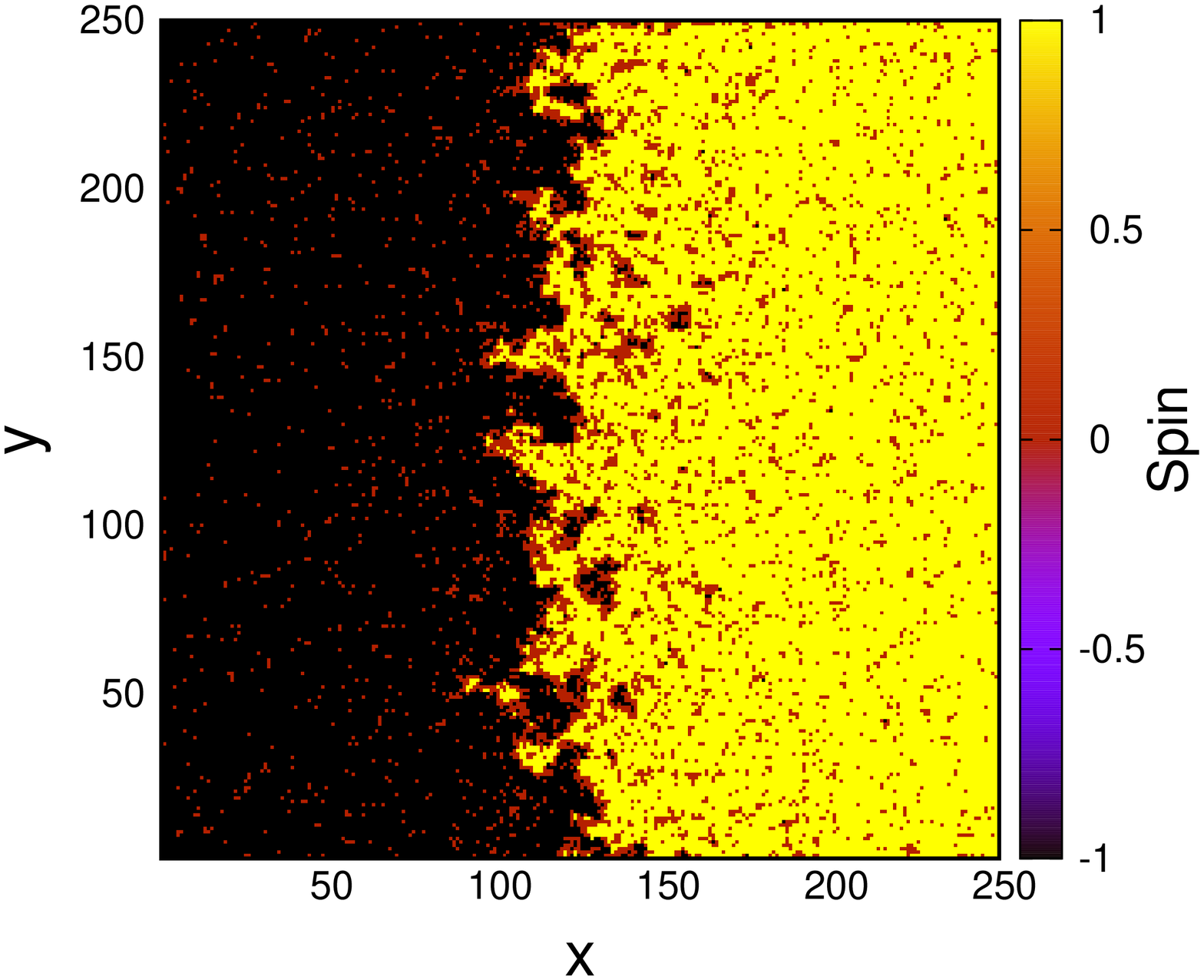}
		\subcaption{}
	\end{subfigure}
	\caption{Snapshots are taken \textbf{at the time of reversal} in presence of uniform 
		field and 3 different strengths of gradients of field (a) uniform field 
		$h= -0.8$, $\tau= 12$ MCSS, (b) $h_l= -0.8$, $h_r= -0.3$,  $G_h= 0.002$, 
		$\tau= 21$ MCSS, (c) $h_l= -0.8$, $h_r= -0.1$,  $G_h= 0.0028$, $\tau= 28$ 
		MCSS, (d) $h_l= -0.8$, $h_r= +0.3$, $G_h= 0.0044$, $\tau= 71$ MCSS. 
		Anisotropy of the system is $\textbf{D= 1.6}$. Temperature is set to 
		$\textbf{T= 0.8}$.}
	\label{snap_gh}
\end{figure}

Some snapshots of the lattice of an anisotropic ($ D=1.6 $) system are taken \textbf{at the 
reversal time} $\tau$ in presence of uniform field and graded field. 
Under uniform field (fig-\ref{snap_gh}a), all the spins '1', '0' and '-1' are uniformly distributed 
throughout the lattice. As we apply some graded field (fig-\ref{snap_gh}b,c,d), spins on left 
side try to flip first due to stronger field. That is why spin '-1' get accumulated strongly 
on left side with the increase in strength of $G_h$. As a result, an \textbf{interface (or domain wall)} appears between 
the two regions (left region is dominated by spin '-1' whereas the right by spin '1'). 

\begin{figure}[h!]
	\centering
	\begin{subfigure}{0.45\textwidth}
		\includegraphics[angle=-90,width=\textwidth]{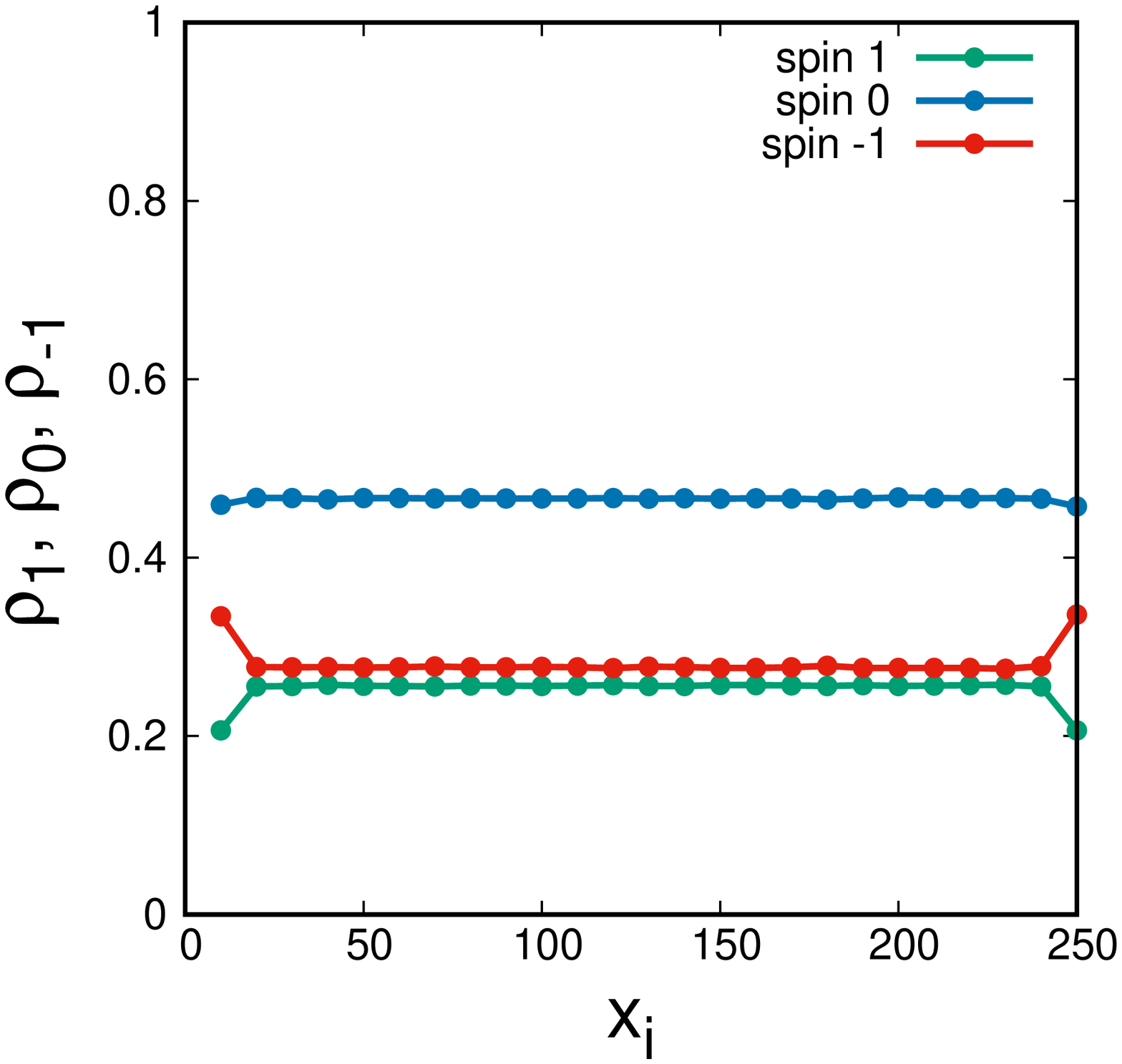}
		\subcaption{}
	\end{subfigure}
	\begin{subfigure}{0.45\textwidth}
		\includegraphics[angle=-90,width=\textwidth]{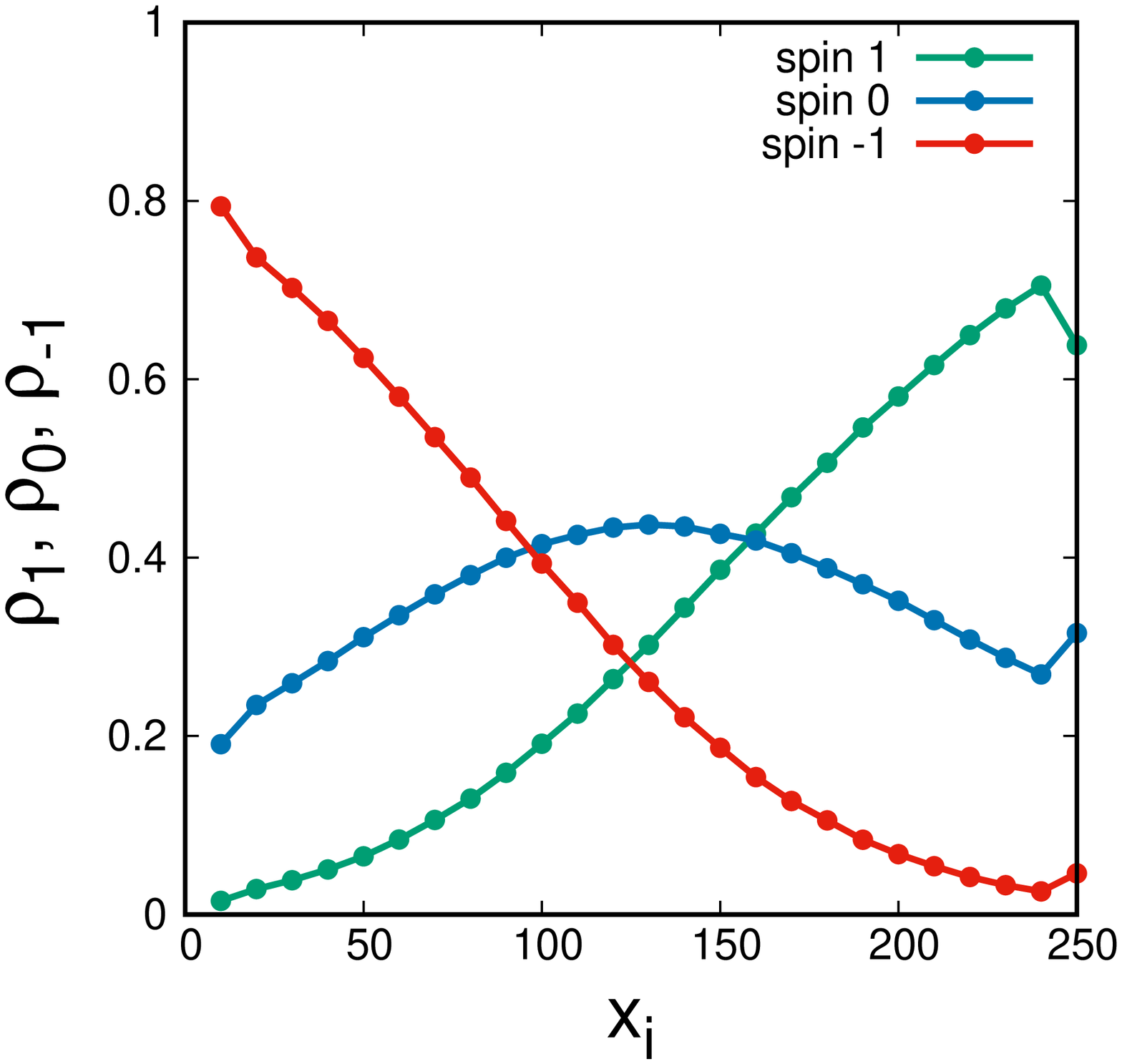}
		\subcaption{}
	\end{subfigure}	
	\begin{subfigure}{0.45\textwidth}
		\includegraphics[angle=-90,width=\textwidth]{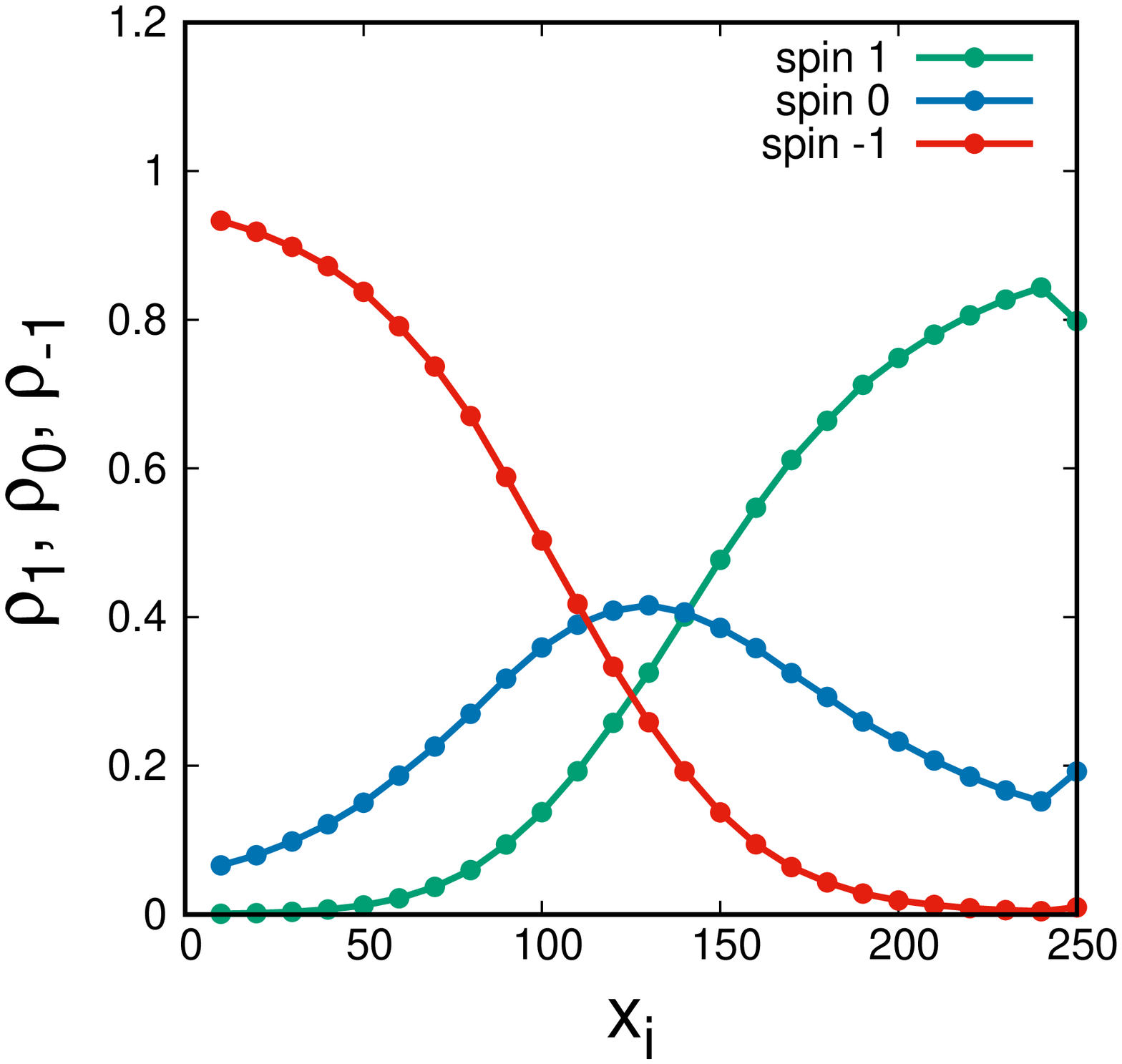}
		\subcaption{}
	\end{subfigure}
	\begin{subfigure}{0.45\textwidth}
		\includegraphics[angle=-90,width=\textwidth]{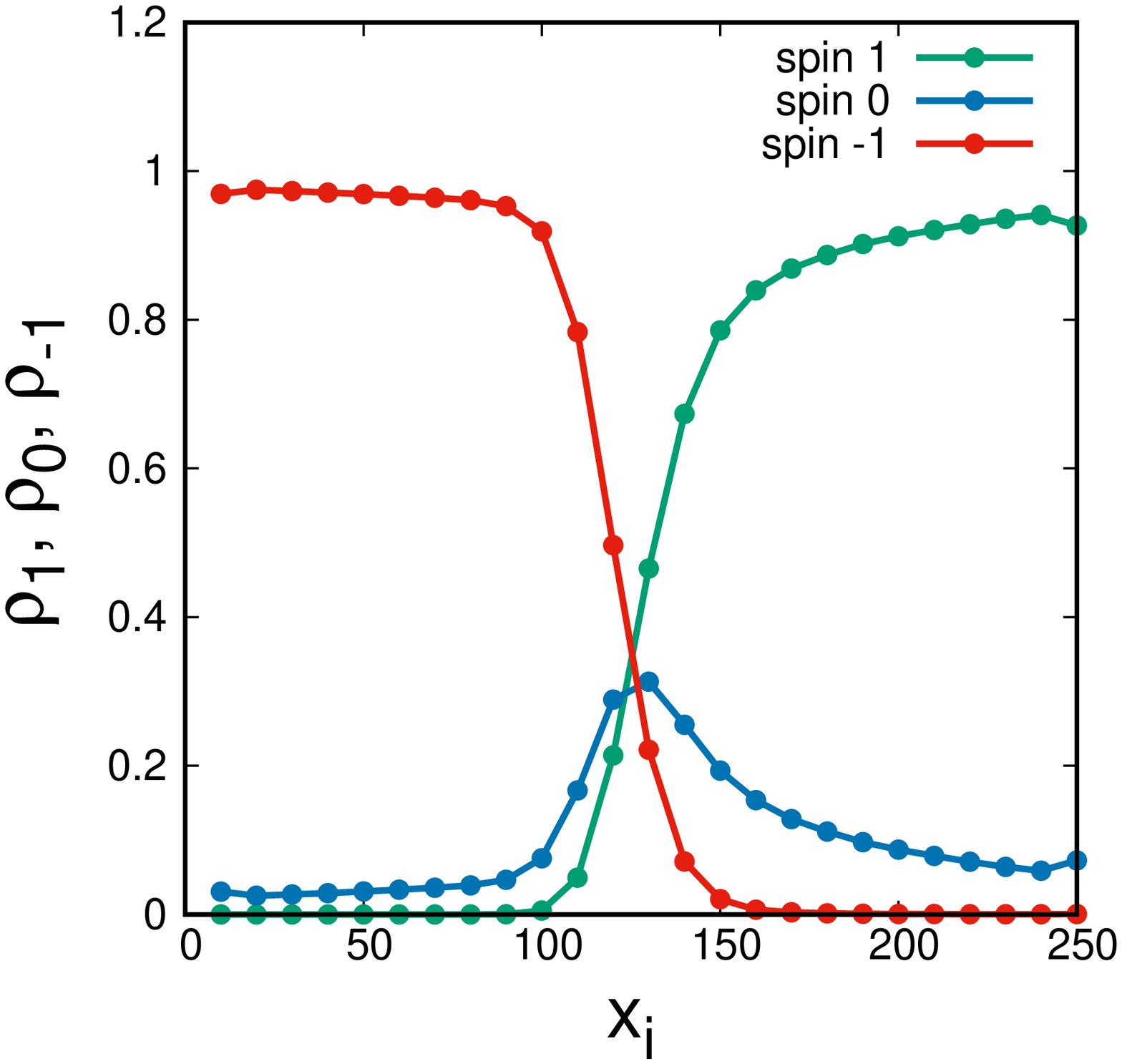}
		\subcaption{}
	\end{subfigure}
	\caption{Variation of density of spin '+1' ($\rho_1$), spin '0' ($\rho_0$), 
		spin '-1' ($\rho_{-1}$) with position along x direction ($x_i$) at the 
		time of reversal in presence of (a) 
		uniform field $h= -0.8$, (b) graded field, $h_l= -0.8$, $h_r= -0.3$,  
		$G_h= 0.002$, (c) graded field, $h_l= -0.8$, $h_r= -0.1$,  $G_h= 0.0028$, (d) graded field, $h_l= -0.8$, $h_r= +0.3$, $G_h= 0.0044$. Anisotropy of the system is $\textbf{D= 1.6}$. Temperature is set to 
		$\textbf{T= 0.8}$.}
	\label{den_gh}
\end{figure}

\begin{figure}[h!]
	\centering
	\begin{subfigure}{0.45\textwidth}
		\includegraphics[angle=-90,width=\textwidth]{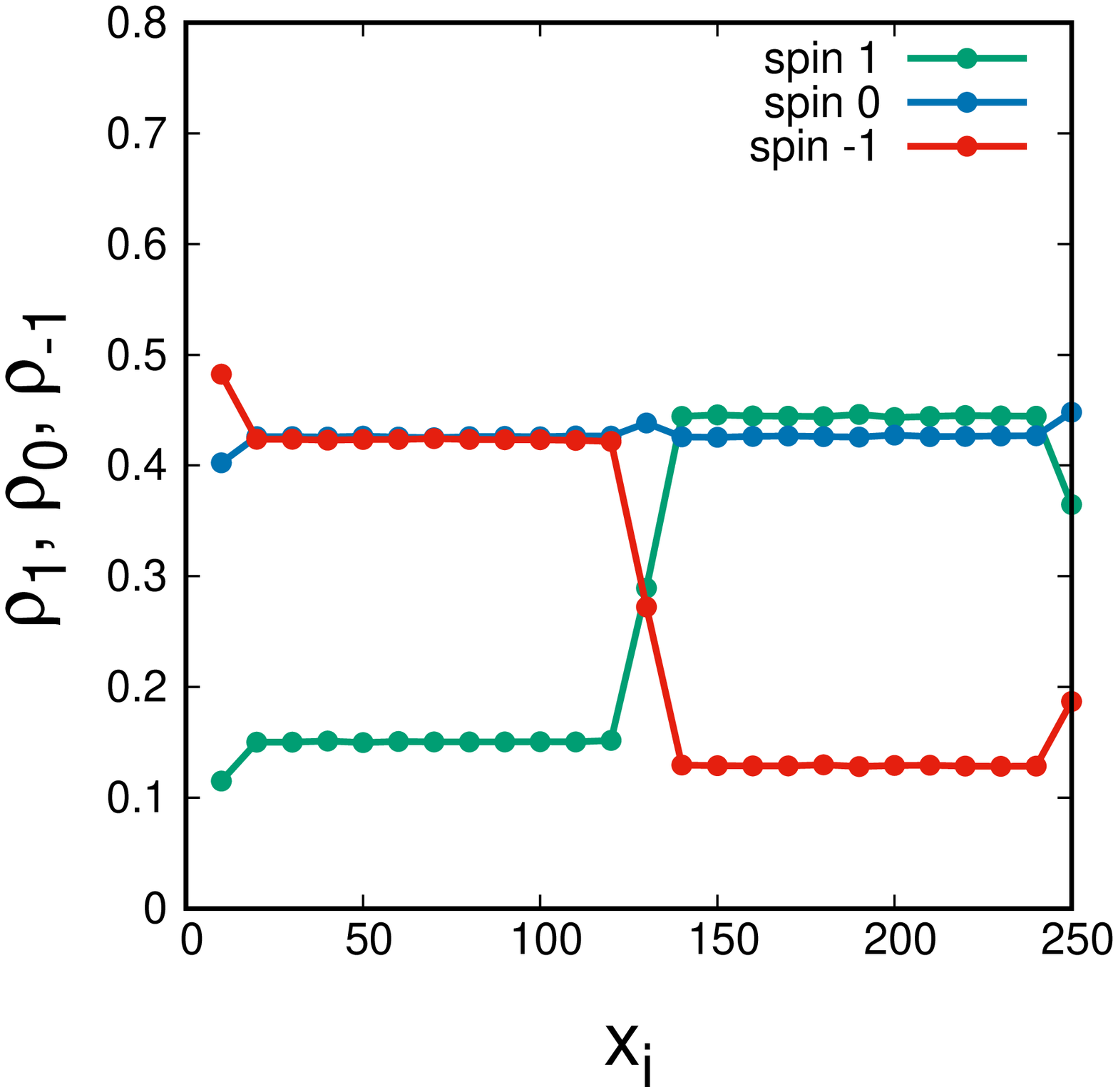}
		\subcaption{}
	\end{subfigure}
	\begin{subfigure}{0.45\textwidth}
		\includegraphics[angle=-90,width=\textwidth]{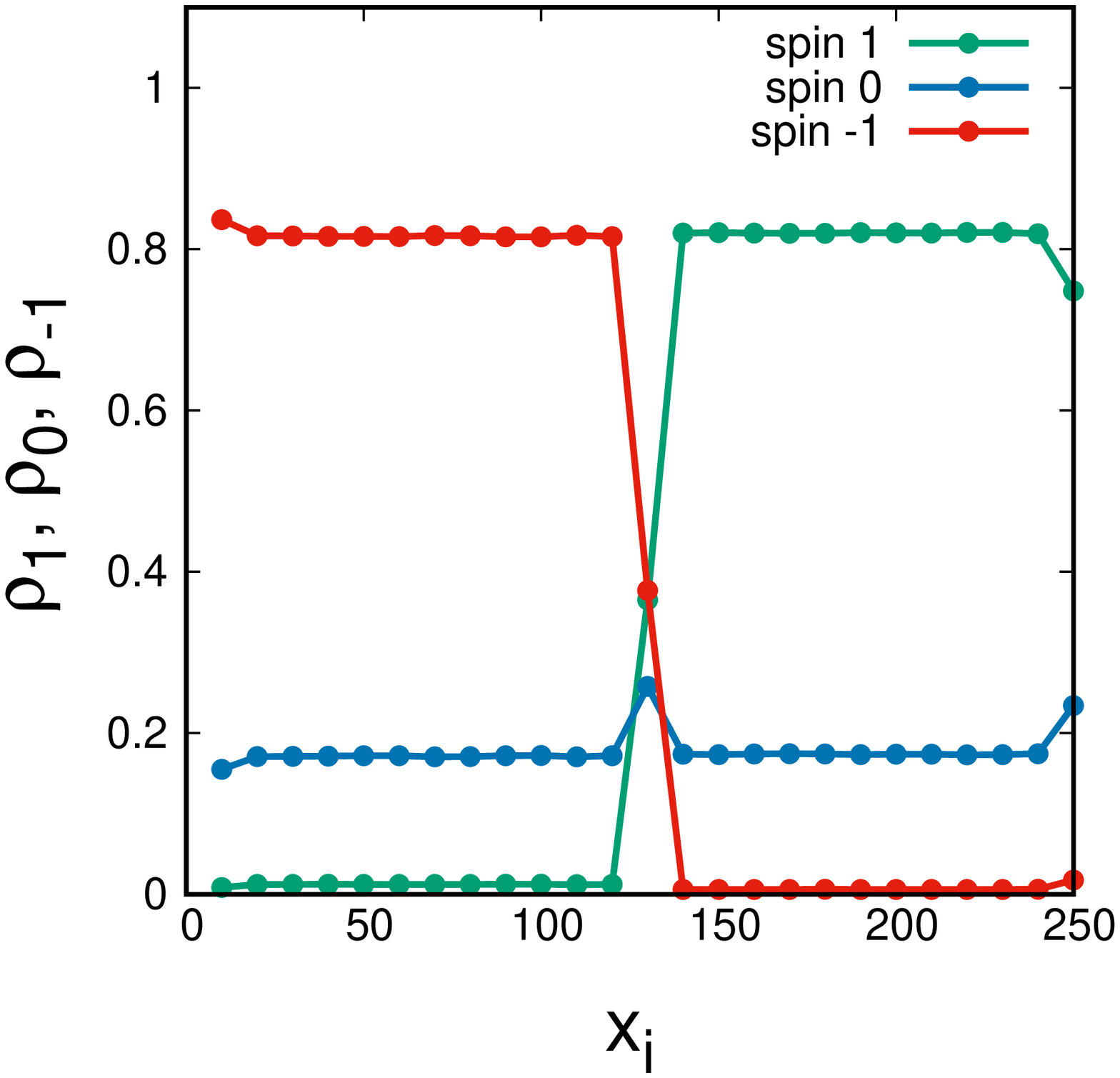}
		\subcaption{}
	\end{subfigure}
	\caption{Variation of density of spin '+1' ($\rho_1$), spin '0' ($\rho_0$), 
		spin '-1' ($\rho_{-1}$) with position along x direction ($x_i$) at the 
		time of reversal in presence of stepped field (a) $h_{sl}= -0.8$, $h_{sr}= -0.6$,  
		$S_h= 0.2$, (b) $h_{sl}= -0.8$, $h_{sr}= -0.2$,  
		$S_h= 0.6$. Anisotropy of the system is $\textbf{D= 1.6}$. Temperature is set to 
		$\textbf{T= 0.8}$.}
	\label{den_sh}
\end{figure}

To have a 
clear idea on spin-dynamics in an anisotropic system to achieve reversal, spatial variation (along x direction only) 
of the density of spin 1,0 and -1 ($\rho_1,\rho_0$ and $\rho_{-1},$) has been checked out. 
To study this, the lattice has been divided into 25 strips (each strip of size $10 \times 250$) 
along x-direction. In each strip $\rho_{-1}$ has been obtained \textbf{at the time of reversal} 
by calculating total number of spin '-1' in a strip and dividing it by the total number of spins 
contained in the strip. And finally the data are averaged over 1000 random samples. $\rho_1$ and 
$\rho_0$ are also calculated in the same fashion. In presence of uniform field (fig-\ref{den_gh}a), 
since the impact of the field on each lattice site is equal, density of all the spins remain
spatially uniform throughout the lattice. And $\rho_0$ is higher than $\rho_1$ or $\rho_{-1}$ 
because of the strength of the anisotropy (positive) present in the system. When applying a graded field 
(fig-\ref{den_gh}b), obviously there is a spatial variation of the applied field throughout the 
lattice. Stronger field on left side compels spins to get flipped from '1' or '0' state to '-1' 
state which rises $\rho_{-1}$ and reduces both $\rho_0$ and $\rho_{-1}$ on that side. Whereas 
on right side, due to the weaker field, most of the spins remain in spin '1' state which results 
in a high $\rho_1$ and low $\rho_0$, $\rho_{-1}$. That is why $\rho_{-1}$ decreases from a higher 
value at left edge to a lower value at right edge. A contrasting nature of $\rho_1$ is observed. 
At the middle position of the lattice (around $x_i \simeq 125$), $\rho_1$ and $\rho_{-1}$ become 
equal. In addition $\rho_0$ bends near the two edges i.e. $\rho_0$ becomes almost same in the 
region of strong or weak field. So actually the effect of anisotropy becomes insensitive to the
strength of the field. As the $G_h$ is increased (fig-\ref{den_gh}c\ref{den_gh}d) by reducing the magnitude of  
field $h_r$, the ability of spin-flip on right side goes down. Then most of the spins from left 
side take part in reversal. As a result, a sharp fall of $\rho_{-1}$ as well as a sharp rise 
of $\rho_1$ near the middle of the lattice is found out. In the same way, such kind of density profile is 
also investigated  for stepped field. When a stepped field of small step difference 
(fig-\ref{den_sh}a) is applied, $\rho_0$ remains almost same as that in case of uniform field. 
$\rho_{-1}$ gets higher at left half side for stronger $h_{sl}$ and lower at right for weaker 
$h_{sr}$. In case of large $S_h$ (fig-\ref{den_sh}b), $\rho_0$ is reduced because most of the 
spins at left half side participate in reversal since $h_{sr}$ is very weak. So a similar kind 
of behaviour as in case of graded field is also observed here. $\rho_0$ remains  almost spatially 
uniform in spite of the presence of a step difference in field. \textit{Distortion near edges of the 
lattice in each plot is the reflection of the open boundary condition.}

\begin{figure}[h!]
	\centering
	\begin{subfigure}{0.4\textwidth}
		\includegraphics[angle=-90,width=\textwidth]{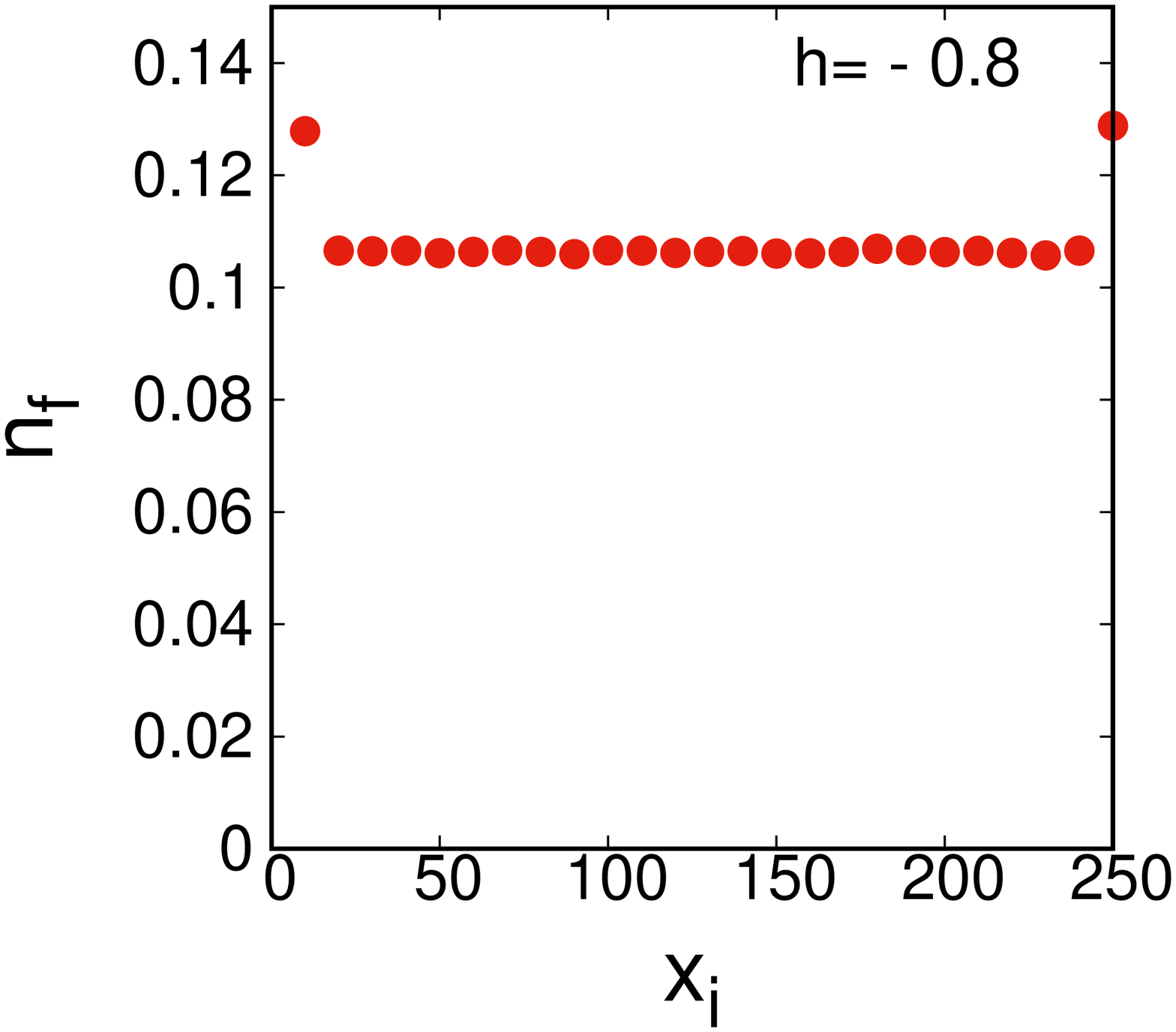}
		\subcaption{}
	\end{subfigure}
	\begin{subfigure}{0.4\textwidth}
		\includegraphics[angle=-90,width=\textwidth]{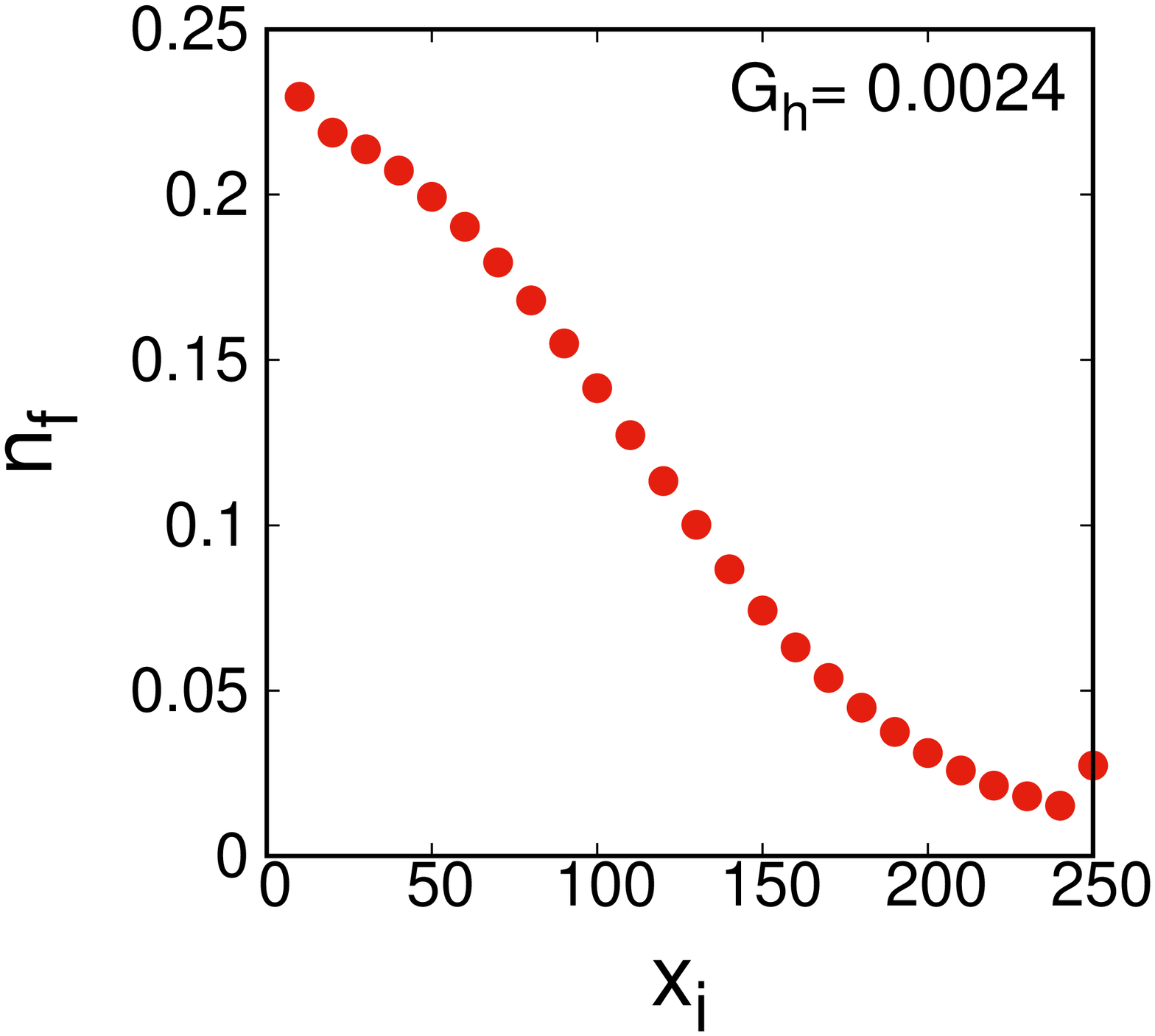}
		\subcaption{}
	\end{subfigure}
	\begin{subfigure}{0.4\textwidth}
		\includegraphics[angle=-90,width=\textwidth]{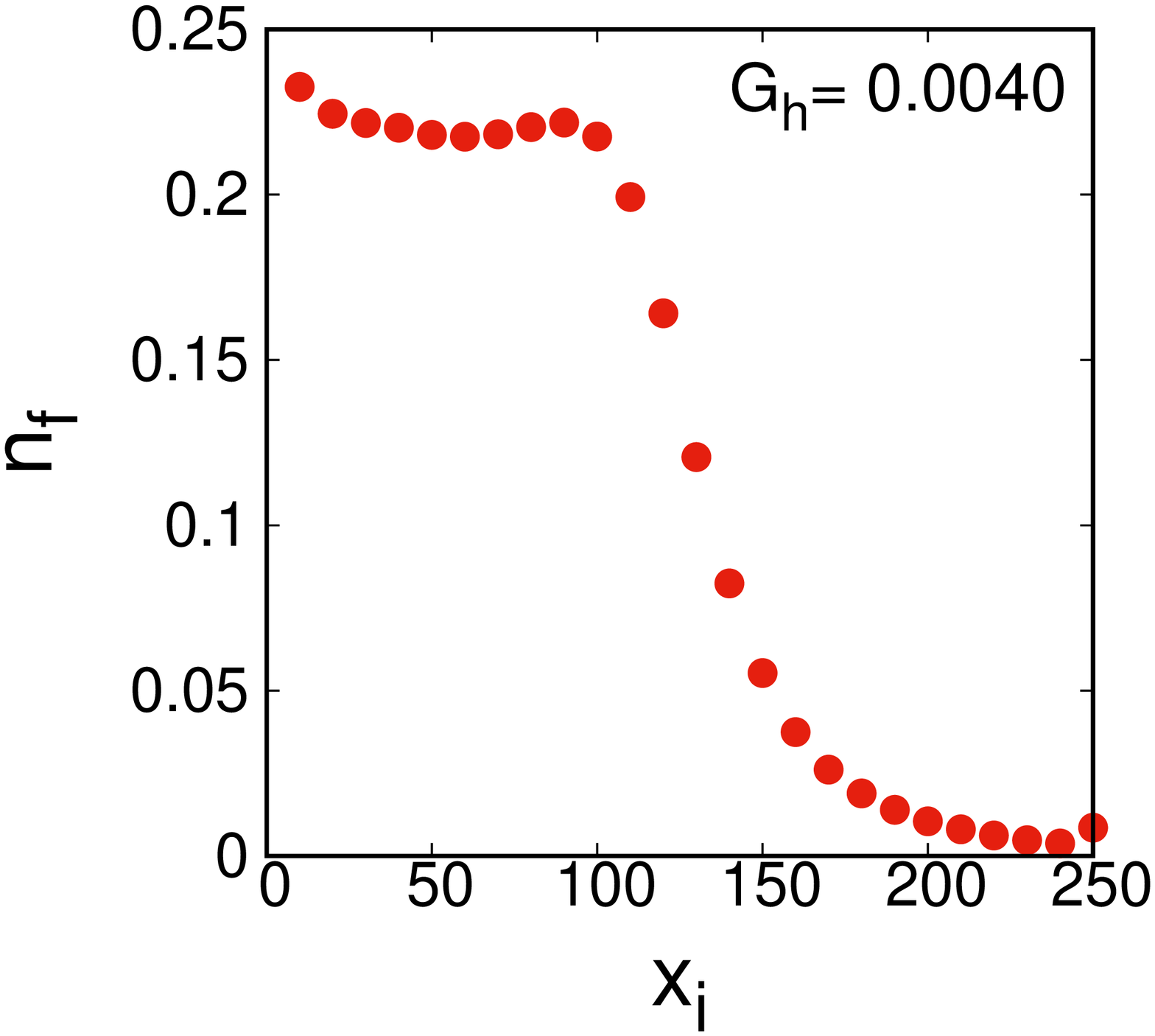}
		\subcaption{}
	\end{subfigure}	
	\begin{subfigure}{0.4\textwidth}
		\includegraphics[angle=-90,width=\textwidth]{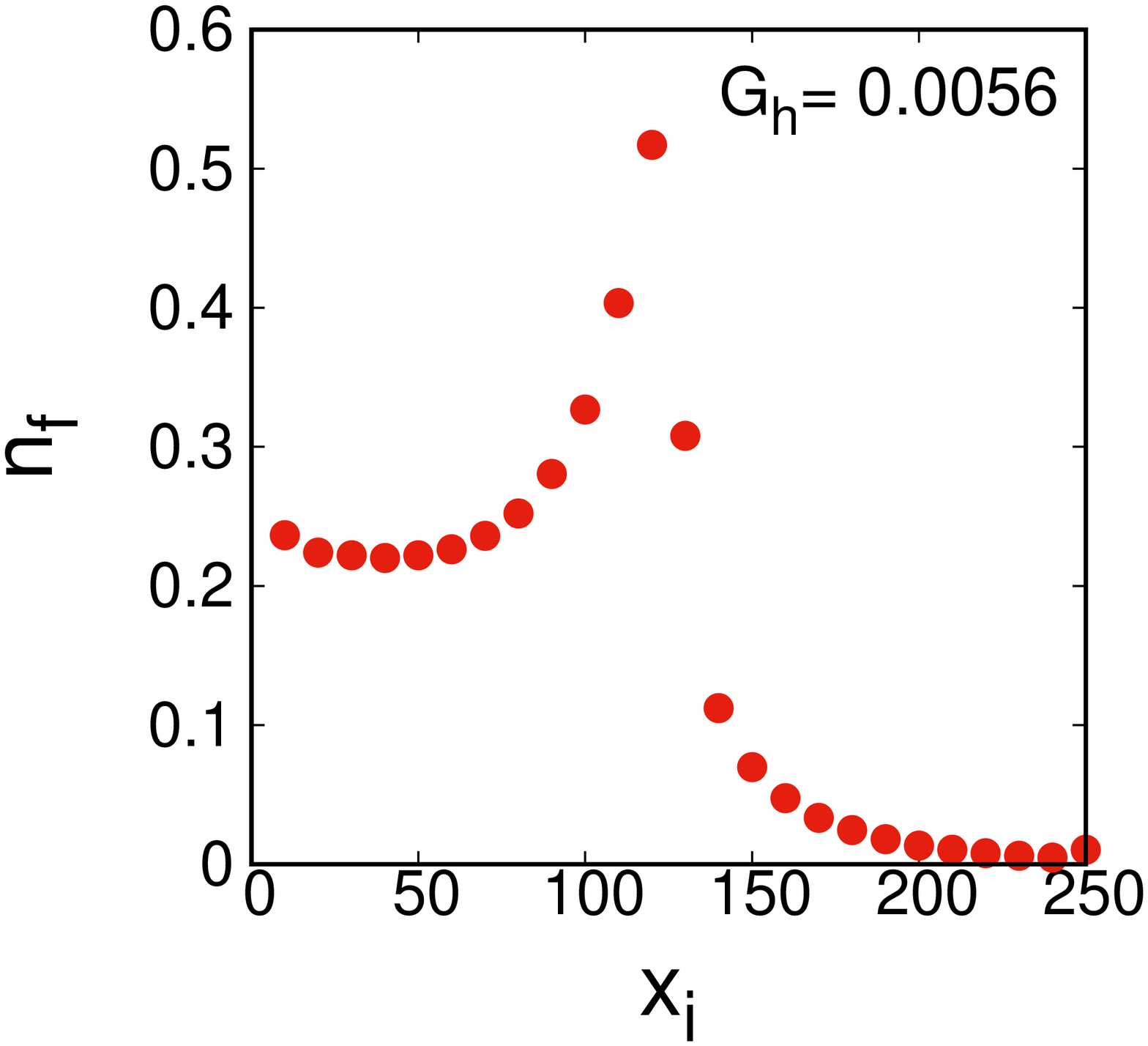}
		\subcaption{}
	\end{subfigure}
	\caption{Spatial variation of number of spin flip (considering 
		flipping from '+1' to '-1' only) upto reversal time per lattice site $n_f$ for the system
		in presence of uniform and graded field. (a) uniform field $h= -0.8$, (b) graded field $h_l= -0.8$, $h_r= -0.2$,  
		$G_h= 0.0024$, (c) graded field $h_l= -0.8$, $h_r= 0.2$,  $G_h= 0.0040$, (d) graded field $h_l= -0.8$, 
		$h_r= 0.6$, $G_h= 0.0056$. Anisotropy of the system is $\textbf{D=1.6}$. 
		Temperature is set to $\textbf{T= 0.8}$.}
	\label{flip_gh}
\end{figure}

\begin{figure}[h!]
	\centering
	\begin{subfigure}{0.4\textwidth}
		\includegraphics[angle=-90,width=\textwidth]{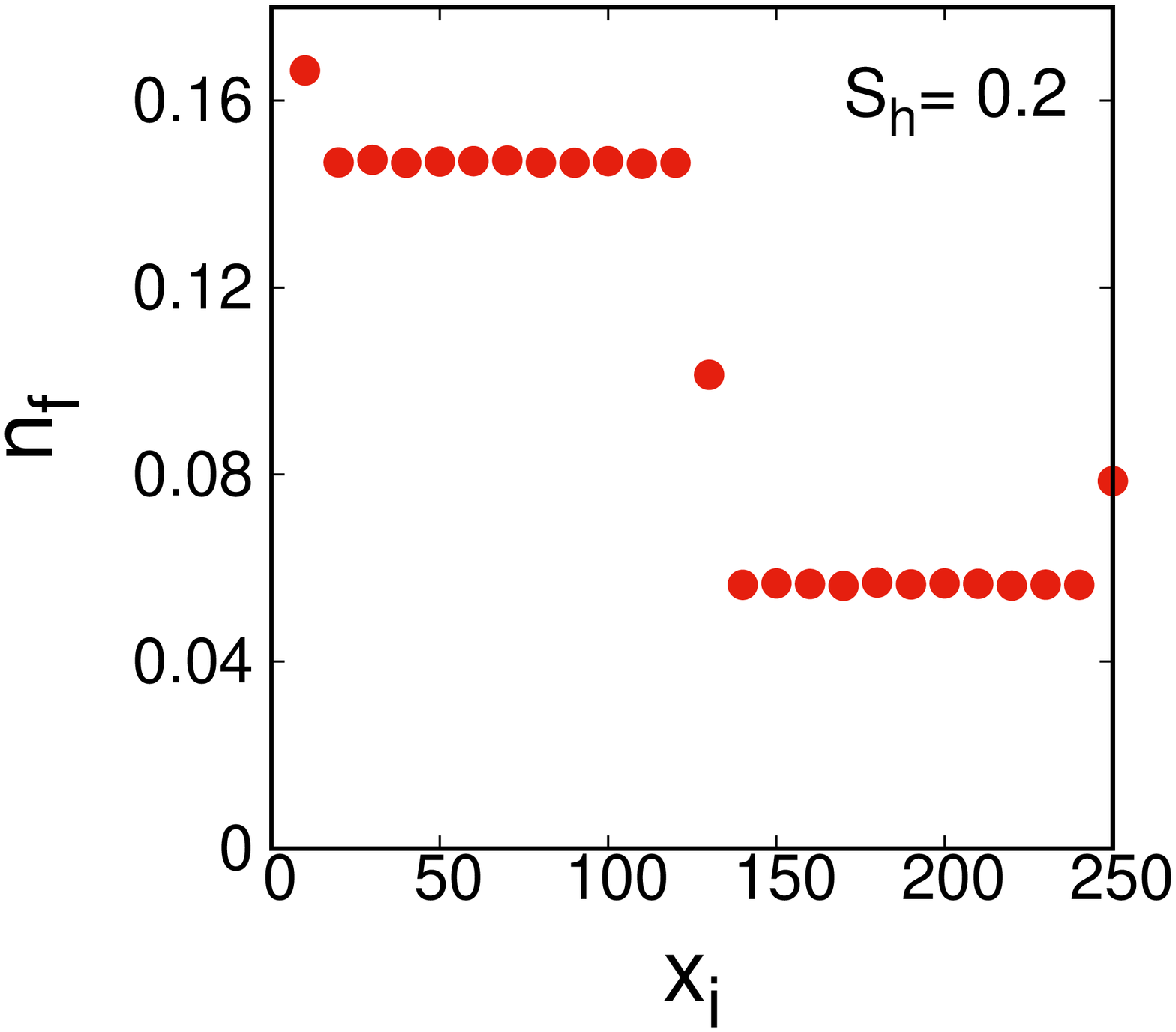}
		\subcaption{}
	\end{subfigure}	
	\begin{subfigure}{0.4\textwidth}
		\includegraphics[angle=-90,width=\textwidth]{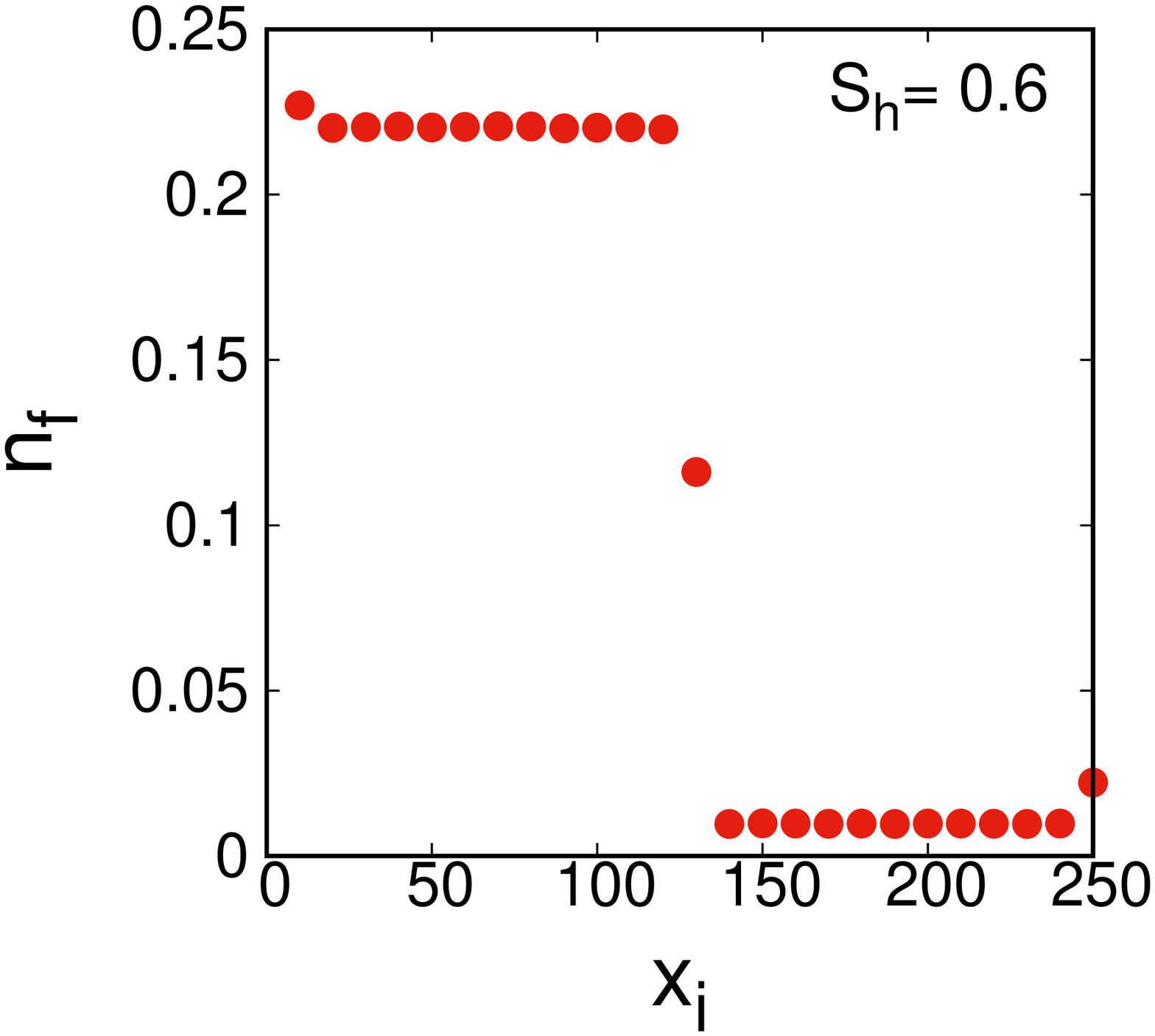}
		\subcaption{}
	\end{subfigure}
	\caption{Spatial variation of number of spin flip (considering flipping from '+1'
		 to '-1' only) upto reversal time per lattice site $n_f$ for the system in stepped field (a) $h_{sl}= -0.8$, $h_{sr}= -0.6$, $S_h= 0.2$, 
		 (b) $h_{sl}= -0.8$, $h_{sr}= -0.2$, $S_h= 0.6$.Anisotropy of the system is $\textbf{D=1.6}$. Temperature is set to 
		 $\textbf{T= 0.8}$.}
	
	\label{flip_sh}
\end{figure}

\begin{figure}[h!]
	\centering
	\begin{subfigure}{0.4\textwidth}
		\includegraphics[angle=-90,width=\textwidth]{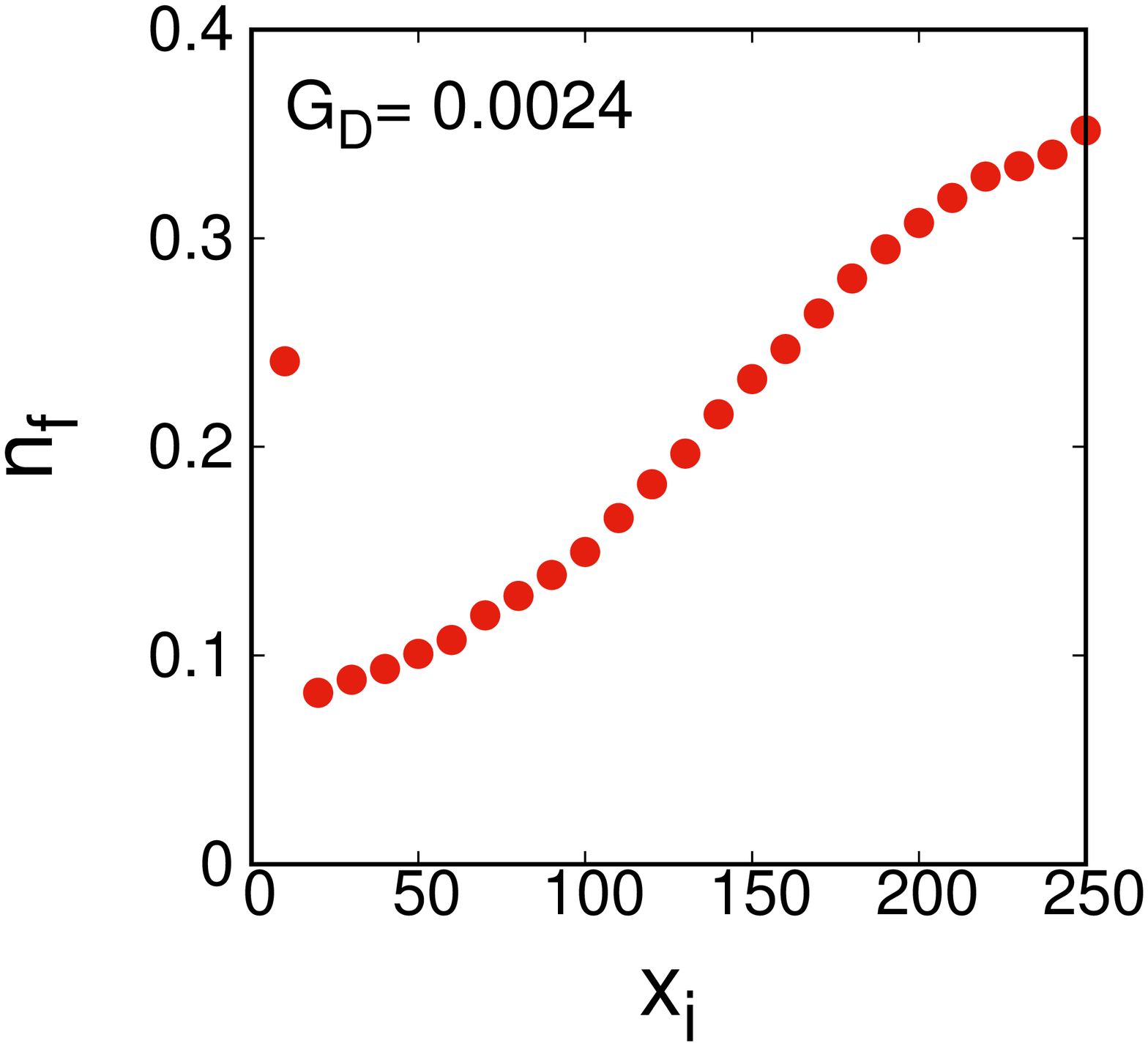}
		\subcaption{}
	\end{subfigure}	
	\begin{subfigure}{0.4\textwidth}
		\includegraphics[angle=-90,width=\textwidth]{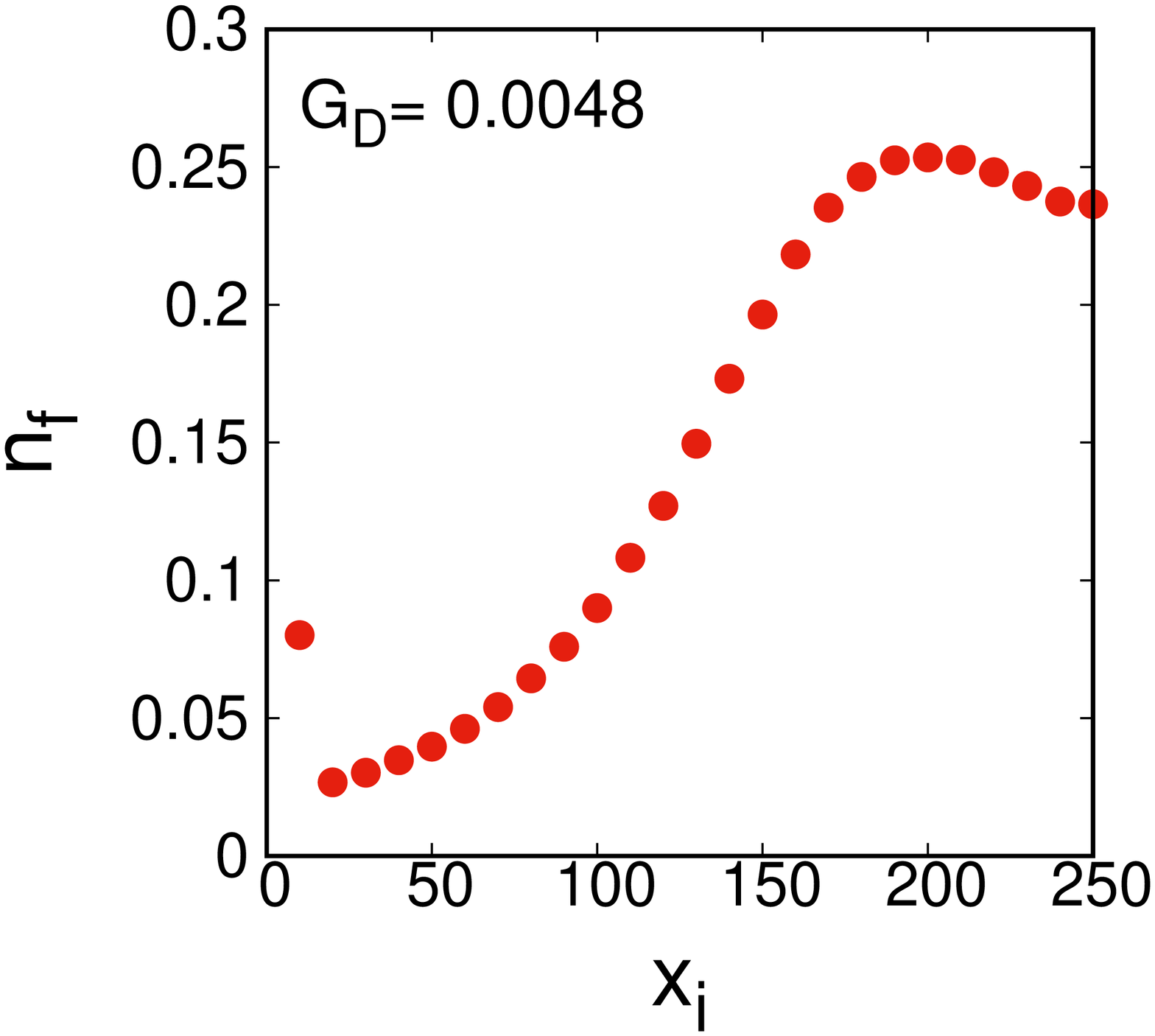}
		\subcaption{}
	\end{subfigure}
	\caption{Spatial variation of number of spin flip (considering flipping from '+1' 
		to '-1' only) upto reversal time per site $n_f$ for graded anisotropic system 
		(a) $D_l= 0.4$, $D_r= 1.0$, $G_D= 0.0024$, (b) $D_l= 0.4$, $D_r= 1.6$,  $G_D= 0.0048$.  
		Applied field is fixed at $\textbf{h= -0.8}$. Temperature is $\textbf{T= 0.8}$.}
	\label{flip_gd}
\end{figure}

\begin{figure}[h!]
	\centering
	\begin{subfigure}{0.4\textwidth}
		\includegraphics[angle=-90, width=\textwidth]{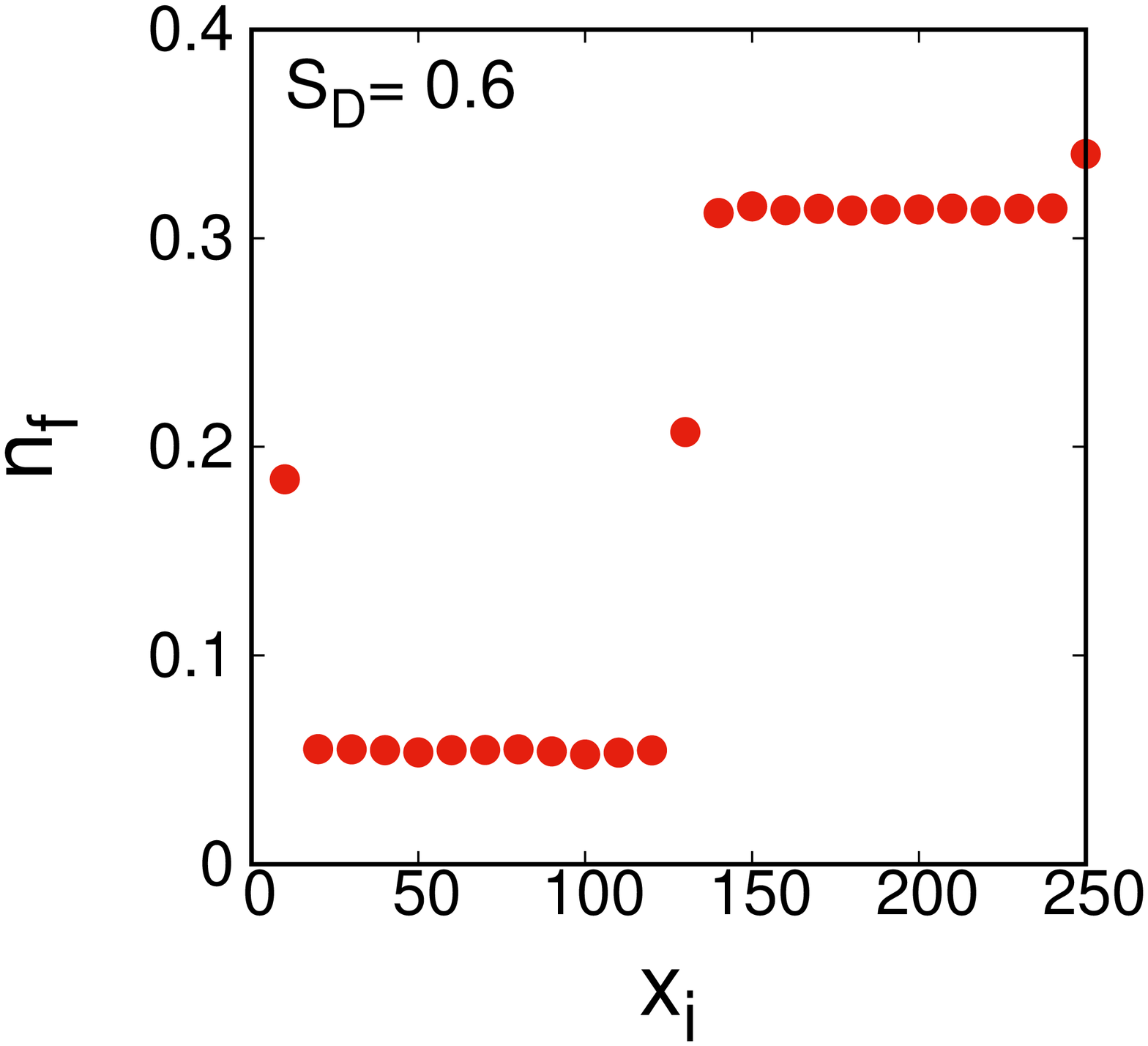}
		\subcaption{}
	\end{subfigure}	
	\begin{subfigure}{0.4\textwidth} 
		\includegraphics[angle=-90,width=\textwidth]{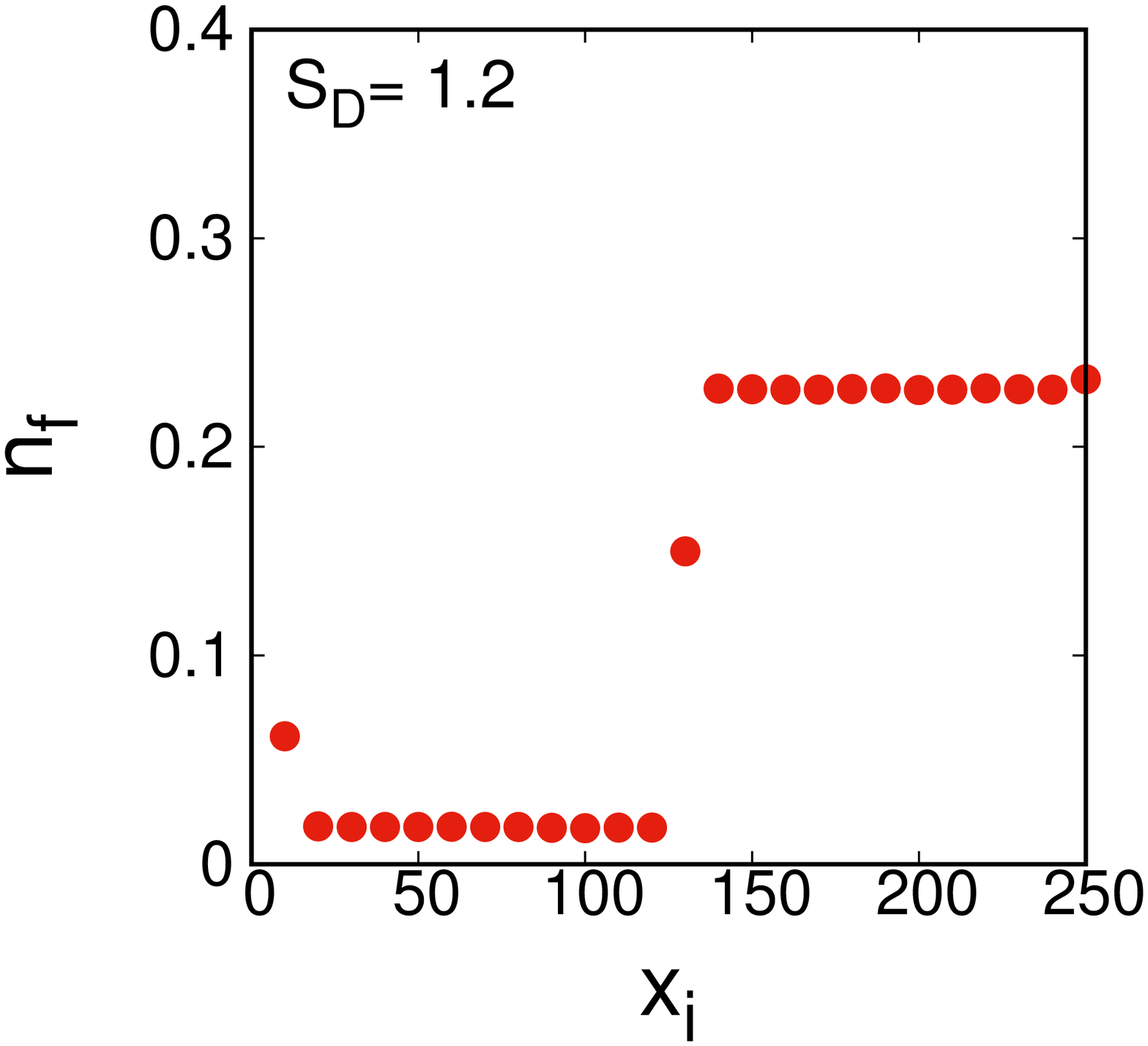}
		\subcaption{}
	\end{subfigure}
	\caption{Spatial variation of number of spin flip (considering 
		flipping from '+1' to '-1' only) upto reversal time per site $n_f$ for the stepped anisotropic system (a) $D_{sl}= 0.4$, 
		$D_{sr}= 1.0$, $S_D= 0.6$, (b) $D_{sl}= 0.4$, $D_{sr}= 1.6$, $S_D= 1.2$. 
		Applied field is fixed at $\textbf{h= -0.8}$. Temperature is set to 
		$\textbf{T= 0.8}$.}
	\label{flip_sd}
\end{figure}

Spatial variation of number of spin flip \textbf{up to the reversal time} per lattice site ($n_f$) is also 
investigated (fig-\ref{flip_gh}) in the same way described above by considering the lattice 
composed of some vertical narrow strips. Total number of spin flip $N_f$ (here considering the 
flipping $\sigma_i^z= +1$ to $-1$ only) up to $\tau$ is calculated in each strip. Then 
dividing it by the total number of sites in each strip we get the $n_f = N_f/ site$.
$n_f$ is also finally averaged over 1000 random samples. In presence of uniform field 
(fig-\ref{flip_gh}a) it remains spatially uniform. 
But under the graded field it becomes higher in the left side compared to the right side (fig-\ref{flip_gh}b,c). 
When the $G_h$ is quite strong, a peak of the $n_f$ is observed near 
the interface (fig-\ref{flip_gh}d). Actually for stronger gradient spins near the left boundary flip very earlier 
and remains stable whereas the flipping possibility near the right boundary is very less due 
to the weak field. In between these two regions, spins try to become stable by balancing the 
effect of two regions which causes higher flipping near the interface. In presence of a stepped 
field (fig-\ref{flip_sh}), $n_f$ is almost uniform having a higher value in the 
left half lattice. In comparison it remains uniform in the right half with a lower 
value since left half of the lattice is under the influence of a stronger field than the 
right half. Same study (fig-\ref{flip_gd}) has been explored over a graded anisotropic system  
(for 2 different strength of $G_D$) in presence of a uniform field $h= -0.8$. The strength of D at the left edge is set to $D_l= 0.4$ and that at the right edge 
is varied ($D_r > D_l$) to vary the $G_D$. Since the strong anisotropy accelerates the spin-flip 
(causes reduction in $\tau$), the number of spin flip rises towards right side. Similarly for 
the system having stepped anisotropy in presence of a uniform field (fig-\ref{flip_sd}), $n_f$ 
remains uniform in the right half lattice with a higher value than the left side due to the 
stronger anisotropy at the right half. The finite discontinuity in the number of spin flips is obvious. \textit{Distortions near edges are present in each plot since we are dealing with open lattice.}

\begin{figure}[h!]
	\begin{subfigure}{0.5\textwidth}
		\includegraphics[angle=-90,width=\textwidth]{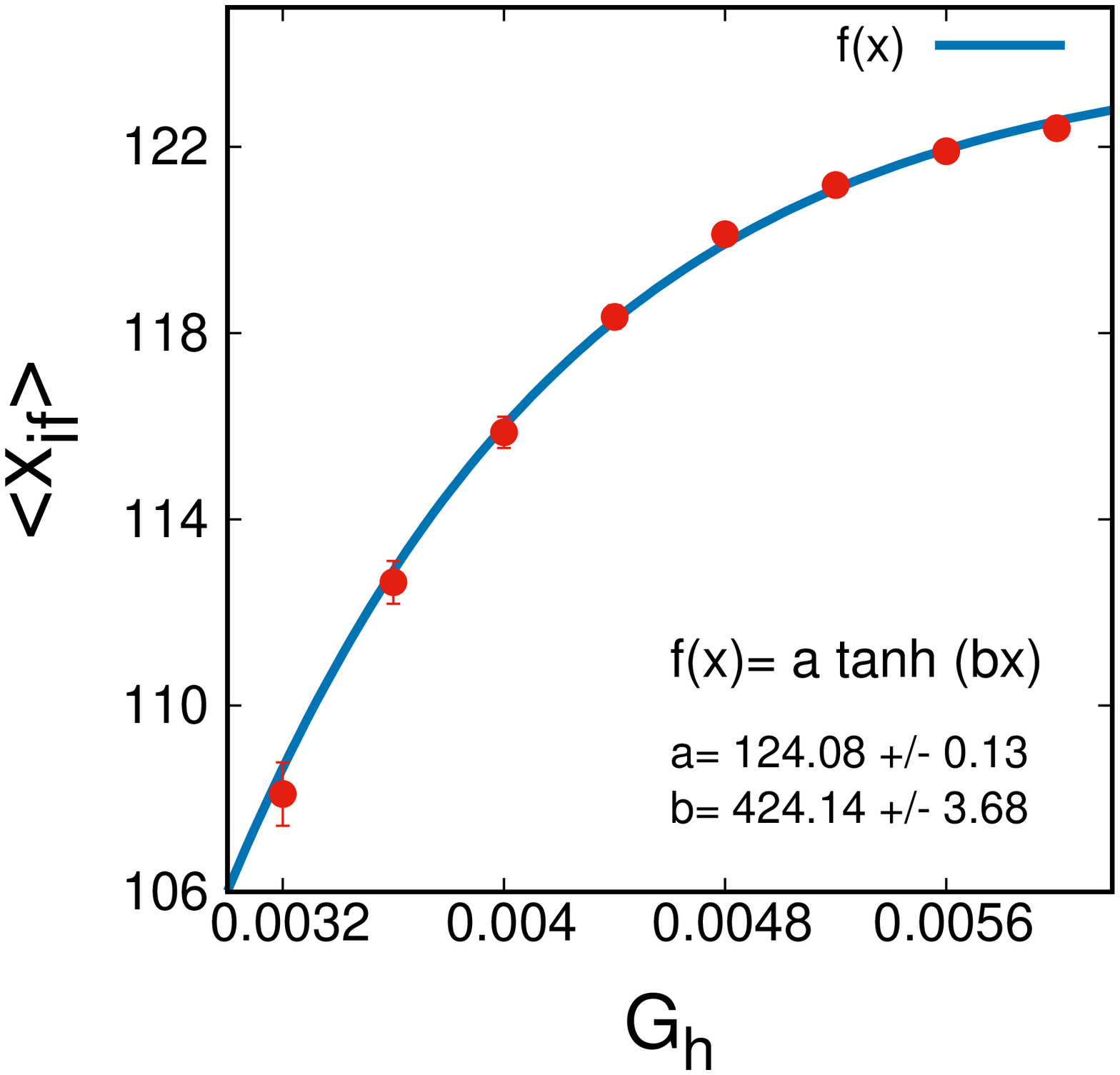}
		\subcaption{}
	\end{subfigure}	
	\begin{subfigure}{0.5\textwidth}
		\includegraphics[angle=-90,width=\textwidth]{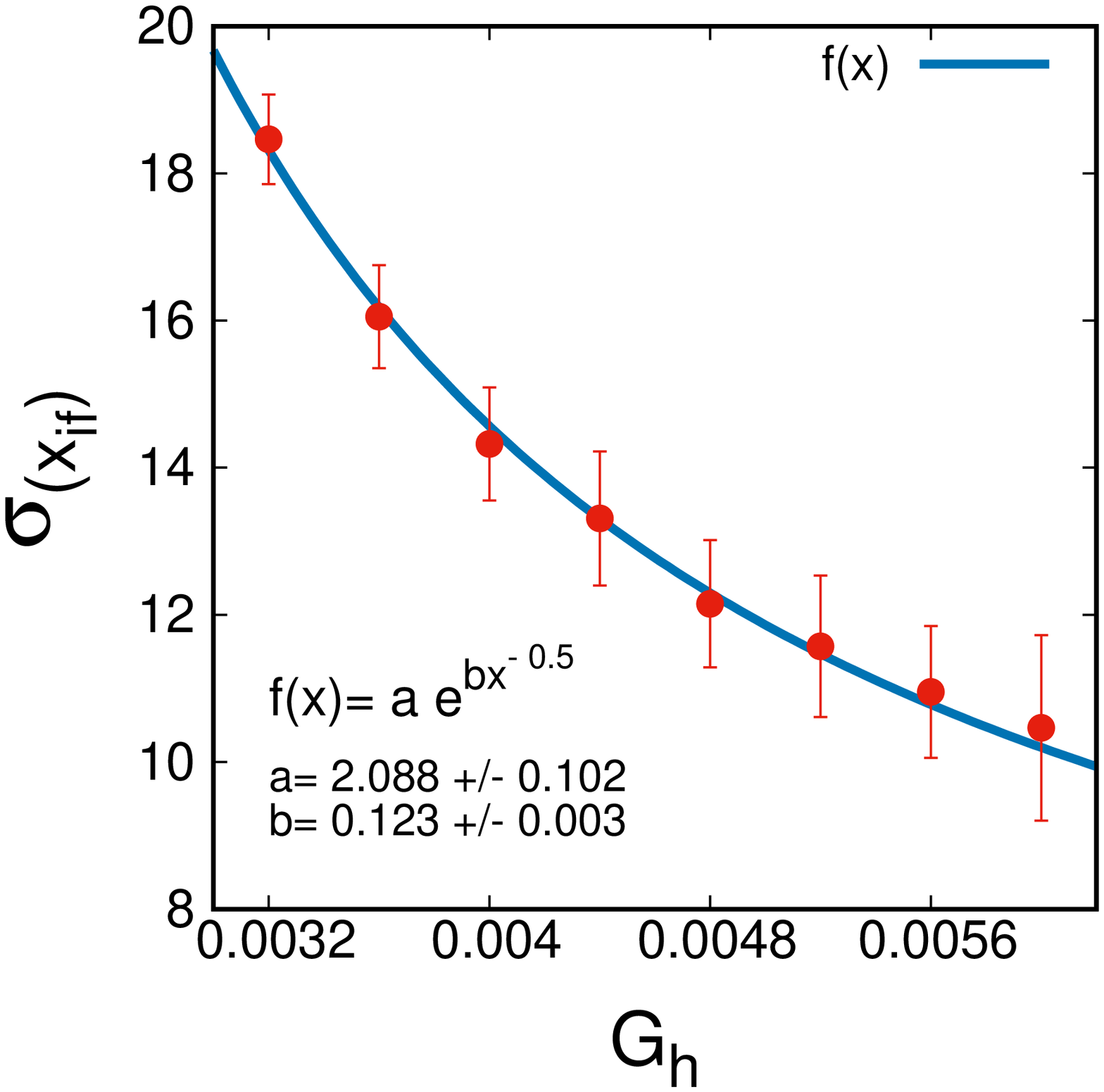}
		\subcaption{}
	\end{subfigure}
	\caption{(a) Variation of average position of interface  at the time of reversal $\langle x_{if} \rangle$ with gradient of field $G_h$ and (b) Variation of 
		roughness $\sigma_{(x_{if})}$ of interface with the $G_h$. Although the fitting parameter 'b' is positive here, due to the factor $x^{-0.5}$ in the exponential, roughness is decreasing as $G_h$ increases. Anisotropy is  
		$\textbf{D=1.6}$. Temperature is set to $\textbf{T= 0.8}$. To vary $G_h$, keeping $h_l= -0.8$ fixed, $h_r$ is varied from 0.0 to +0.7.Errors are estimated using Block averaging method by taking the standard deviation of the meanvalue of 100 blocks each containing 10 datapoints.}
	\label{if_gh}
\end{figure}

\begin{figure}[h!]
	\begin{subfigure}{0.5\textwidth}
		\includegraphics[angle=-90,width=\textwidth]{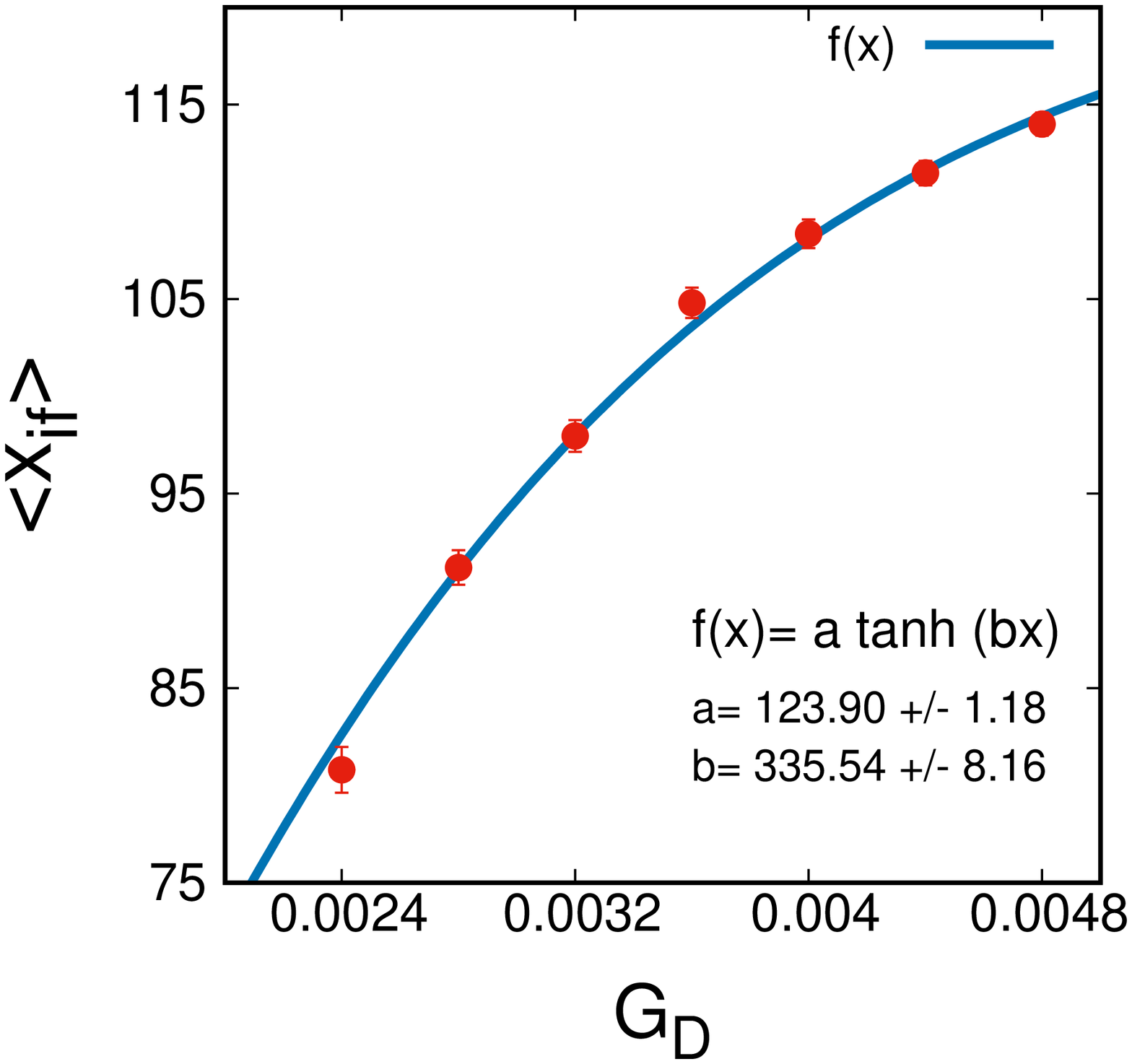}
		\subcaption{}
	\end{subfigure}	
	\begin{subfigure}{0.5\textwidth}
		\includegraphics[angle=-90,width=\textwidth]{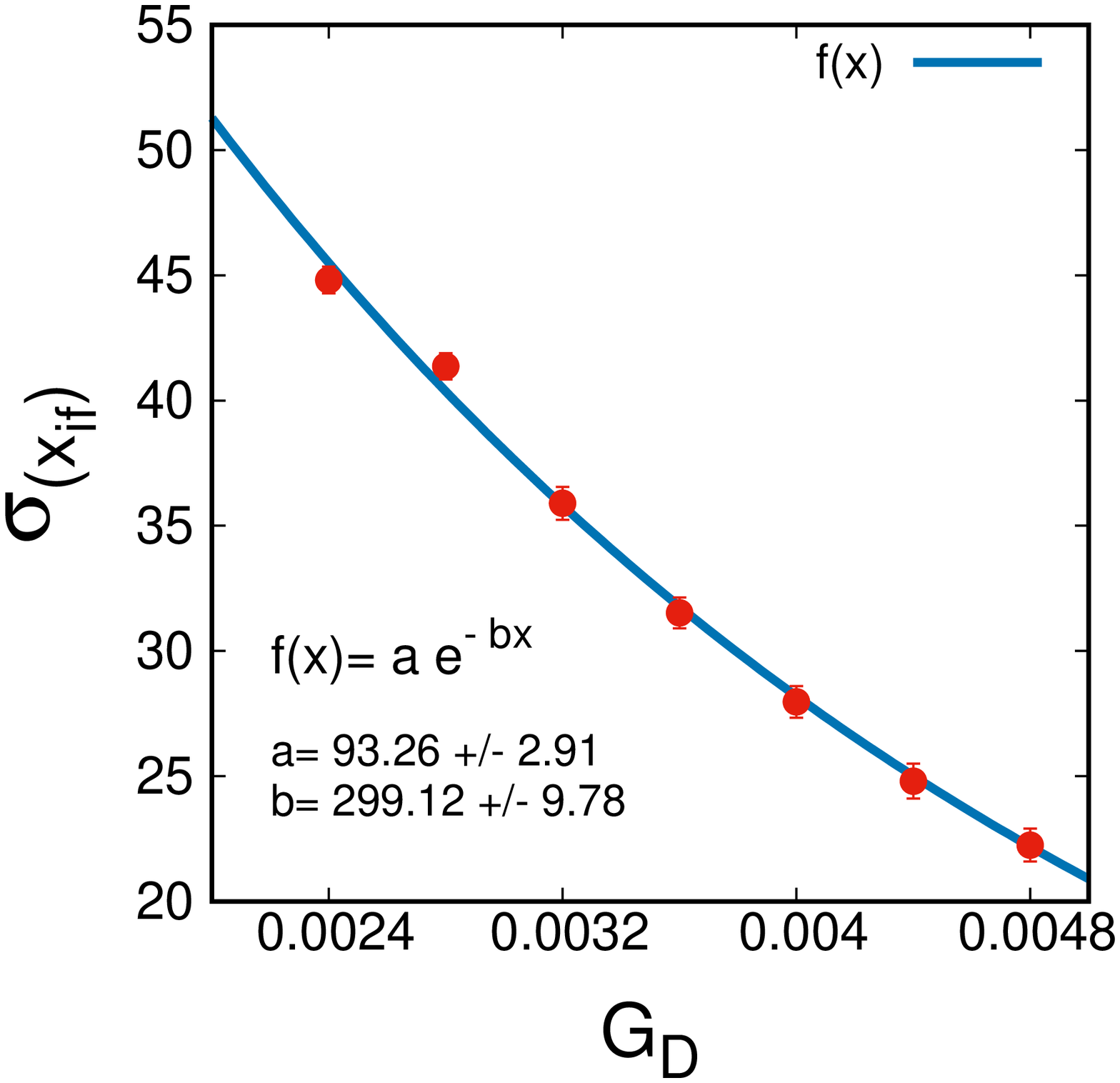}
		\subcaption{}
	\end{subfigure}
	\caption{(a) Variation of average position of interface at the time of reversal $\langle x_{if}\rangle$  with gradient of anisotropy $G_D$ and (b) Variation 
		of roughness of interface $\sigma_{(x_{if})}$ with $G_D$. To vary $G_D$, keeping $D_l= 1.6$ fixed $D_r$ is varied such that $D_r < D_l$. Applied field is kept fixed at 
		$\textbf{h= -0.8}$. Temperature is set to $\textbf{T= 0.8}$.Errors are estimated using Block averaging method by taking the standard deviation of the meanvalue of 100 blocks each containing 10 datapoints.}
	\label{if_gd}
\end{figure}

Lets study about the behaviour of the interface (or domain wall) appearing under graded field or graded 
anisotropy. For the system possessing uniform anisotropy under a particular gradient of field, 
average position of the interface is determined. 
From the snapshots (fig-\ref{if_gh}), clearly at the time of reversal, the left side of the 
lattice is dominated 
by the spin '-1'  and the right side by spin '+1'. Now to determine the position 
of the interface, position of each point on the interface should be found out first. To do so, 
we have 
checked the 10 nearest neighbour spins on both sides of each lattice site. The site for which
the number of spin '-1' on left side is almost equal to the number of spin '+1' on the right side 
is considered as a point on the interface. Thus all the 250 points are found out for whole 
lattice and averaged to get the average position of the interface $\langle x_{if} \rangle$. Then it 
is obtained for 1000 samples and averaged to get final $\langle x_{if} \rangle$ whose variation 
with the $G_h$ follows a tangent hyperbolic function $\langle x_{if} \rangle \simeq 124.08\;{\rm tanh}\; 
(424.14 \;G_h)$ (fig-\ref{if_gh}a). As the $G_h$ is increased, the interface shift towards right 
and finally get fixed near the middle of the lattice. Stronger 
gradient reduces the roughness of the interface and follows the exponential curve 
$\sigma_{(x_{if})} \sim e^{0.12\;G_h^{\;-0.5}}$ (fig-\ref{if_gh}b). Roughness of the interface is 
determined simply by taking the 
standard deviation of the points on the interface and also averaged over 1000 samples. In same 
fashion, above study has been analyzed
for a graded anisotropic system (fig-\ref{if_gd}a) in a uniform field, where also $\langle x_{if} \rangle$ 
varies as a tangent hyperbolic function of $G_D$, 
$\langle x_{if} \rangle \simeq 124\;{\rm tanh}\; (335 \;G_D)$ and the roughness (defined as the 
standard deviation of the interfacial positions) \cite{abyaya} \cite{ranajay} of the interface 
decreases exponentially with the $G_D$ (fig-\ref{if_gd}b), which follows $\sigma_{(x_{if})} \sim 
{\rm exp}(-299\; G_D)$.

\begin{figure}[h!]
	\begin{subfigure}{0.33\textwidth}
		\includegraphics[angle=-90,width=1.1\textwidth]{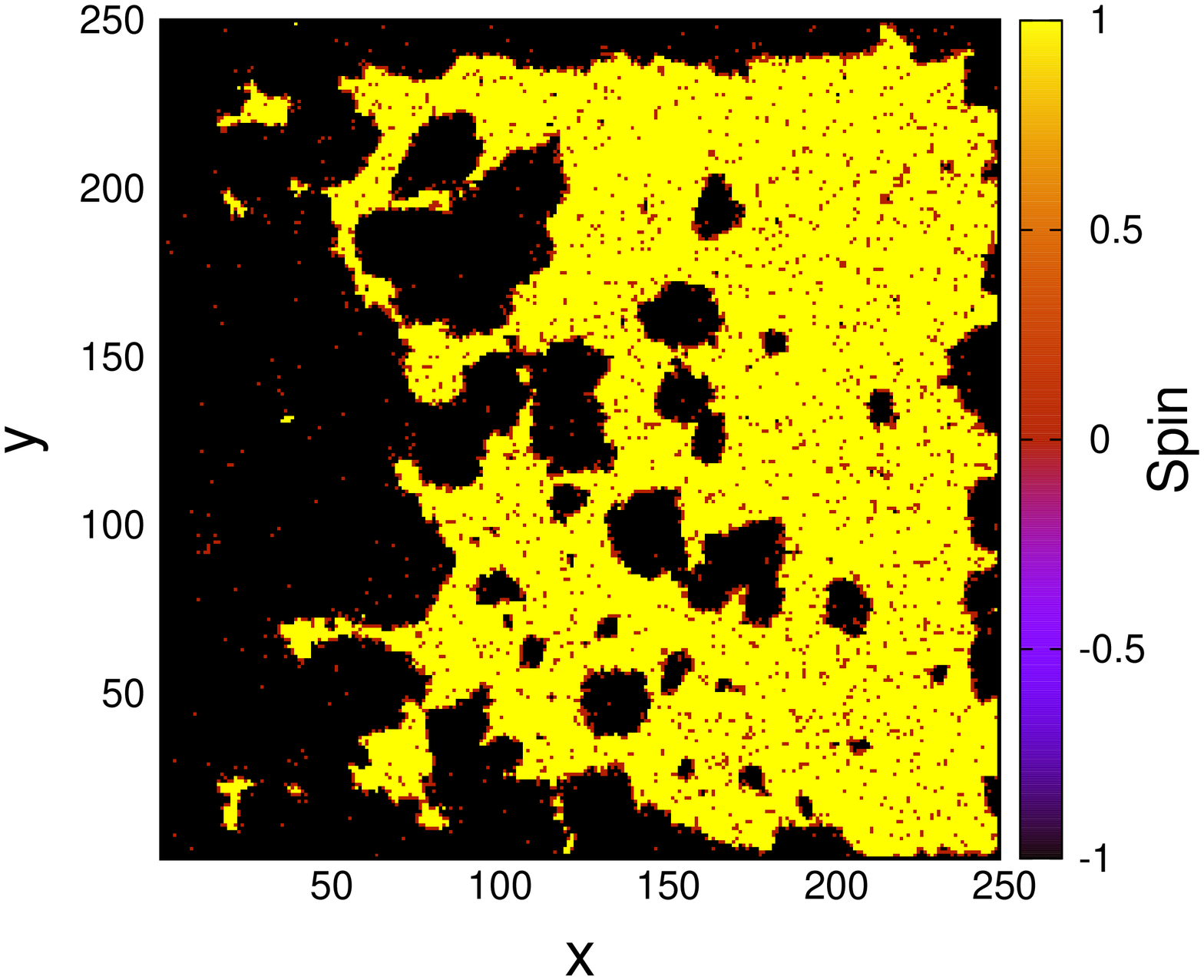}
		\subcaption{}
	\end{subfigure}	
	\begin{subfigure}{0.33\textwidth}
		\includegraphics[angle=-90,width=1.1\textwidth]{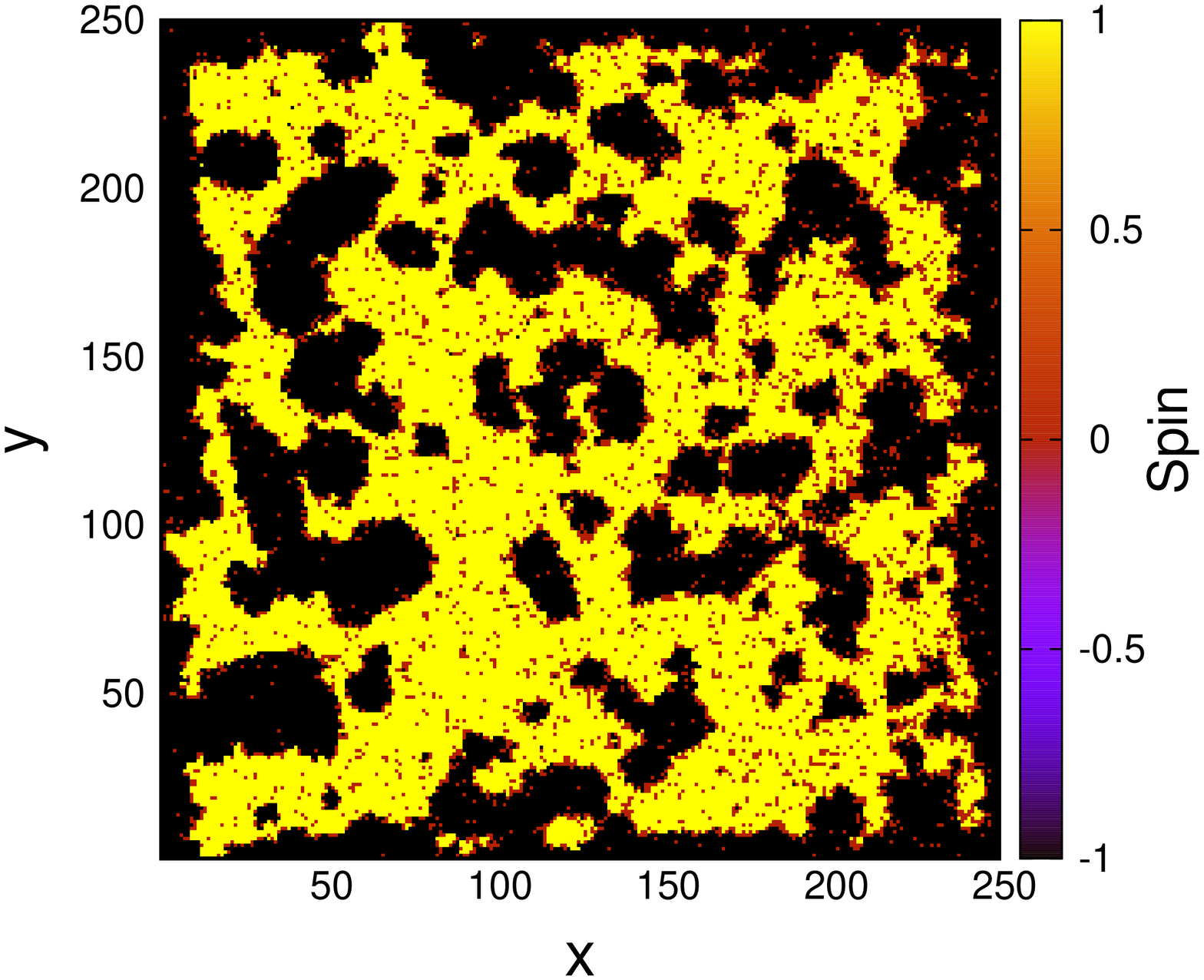}
		\subcaption{}
	\end{subfigure}
	\begin{subfigure}{0.33\textwidth}
		\includegraphics[angle=-90,width=1.1\textwidth]{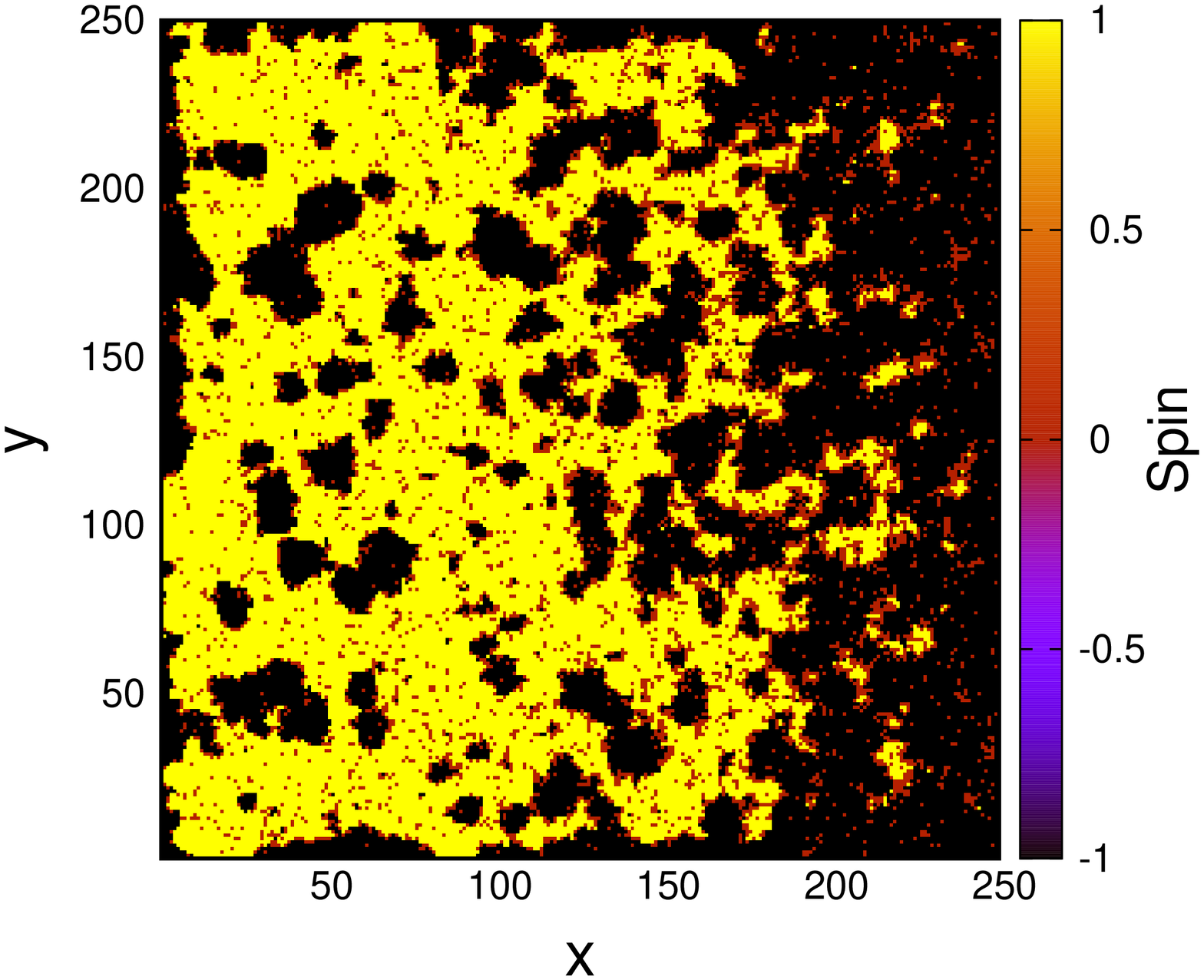}
		\subcaption{}
	\end{subfigure}
	\caption{Snapshots are taken for 3 different graded anisotropic (where anisotropy act competitively with field) system at the time of reversal in presence of a 
		fixed $G_h$. Gradient of field is fixed at $G_h= 0.0016$ ($h_l= -0.8$; $h_r= -0.4$). Keeping $D_l= 0.4$ fixed, $D_r$ is varied to change $G_D$. (a) $D_r= 1.0$,  $G_D= 0.0024$, $\tau= 131$ MCSS, (b) $D_r= 1.35$, 
		$G_D= 0.0038$ (for which $C_F \sim 0.5$ at the $\tau$), $\tau= 85$ 
		MCSS, (c) $D_r= 1.6$, $G_D= 0.0048$, $\tau= 61$ MCSS. Temperature 
		is set to $\textbf{T= 0.8}$.}
	\label{mcomp_snapgh}
\end{figure}

\begin{figure}[h!]
	\begin{center}
		\includegraphics[angle=-90,width=0.5\textwidth]{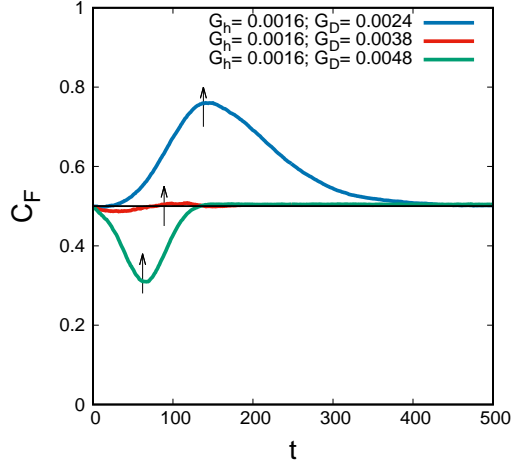}
	\end{center}
	\caption{Evolution of the competition factor $C_F$ (averaged over 100 random samples) with time (MCSS) for 
		the system in fixed $G_h= 0.0016$ and 3 different 
	    $G_D$. Black arrow denotes 
		the reversal time for each and the horizontal black line indicates $C_F= 0.5$. Temperature is set to $\textbf{T= 0.8}$.}
	\label{cftime}
\end{figure}

\begin{figure}[h!]
	\centering
		\includegraphics[angle=-90,width=0.5\textwidth]{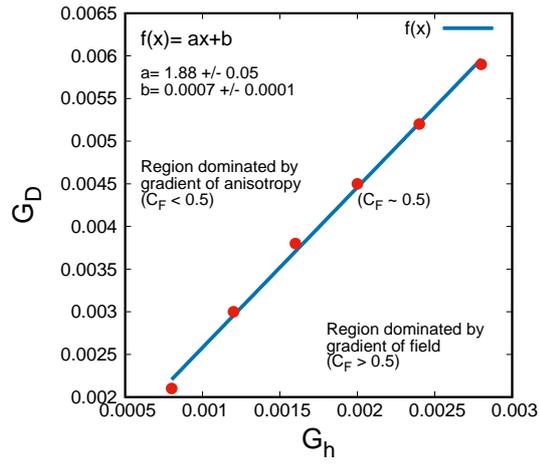}
	\caption{Relation of competitive gradient of anisotropy $G_D$ for 
		which $C_F \sim 0.5$ with the gradient of field $G_h$.}
	\label{mcomp_gh}
\end{figure}

\begin{figure}[h!]
	\begin{subfigure}{0.33\textwidth}
		\includegraphics[angle=-90,width=1.1\textwidth]{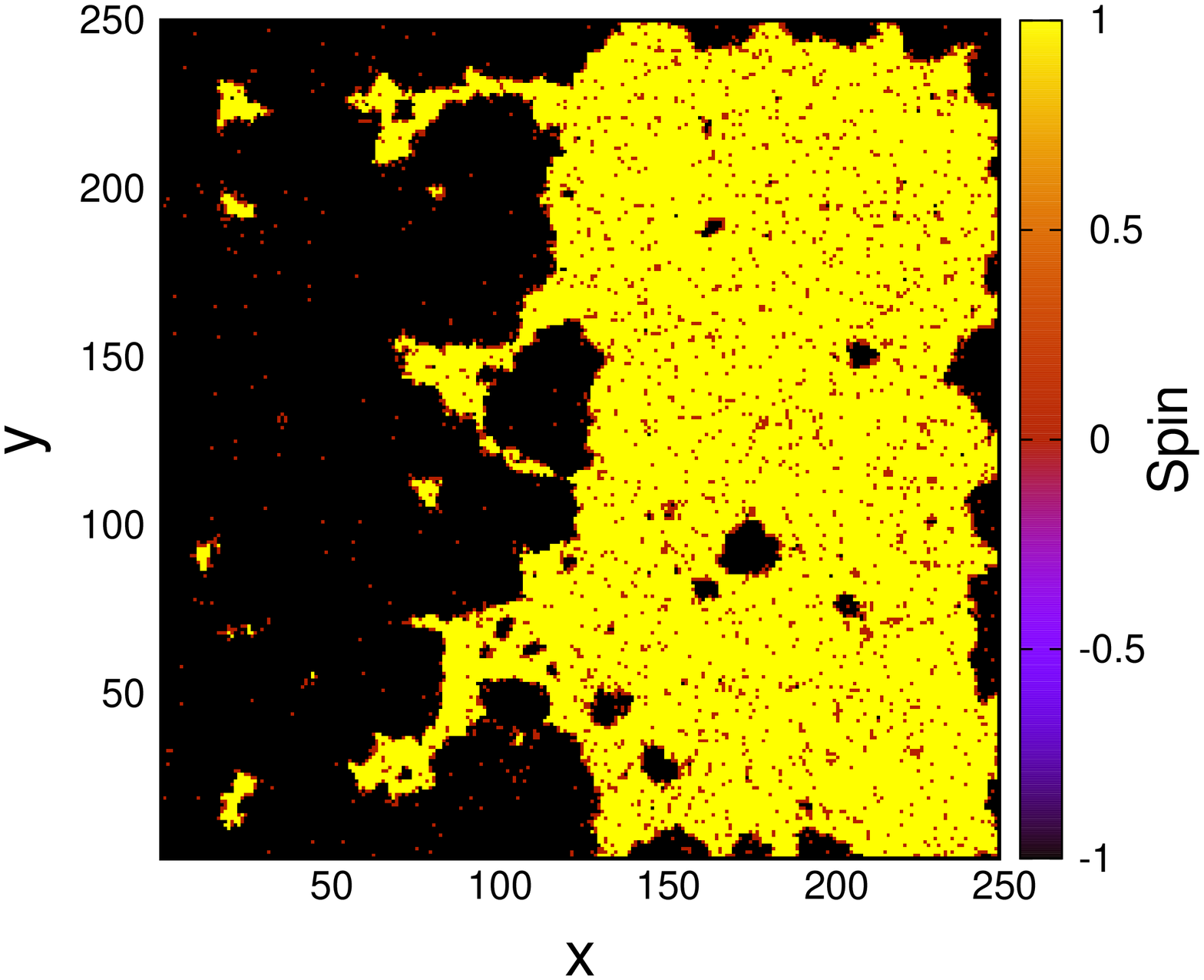}
		\subcaption{}
	\end{subfigure}	
	\begin{subfigure}{0.33\textwidth}
		\includegraphics[angle=-90,width=1.1\textwidth]{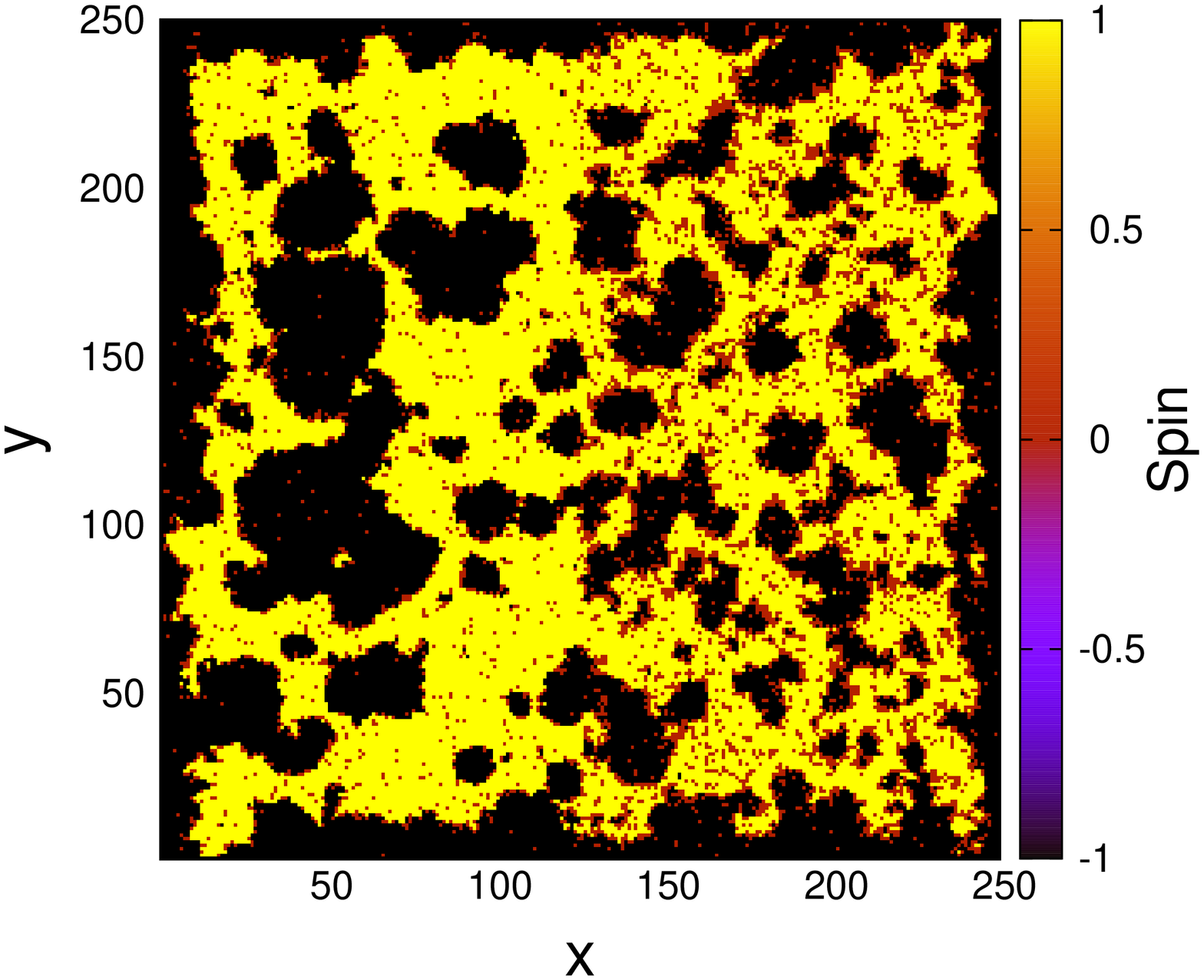}
		\subcaption{}
	\end{subfigure}
	\begin{subfigure}{0.33\textwidth}
		\includegraphics[angle=-90,width=1.1\textwidth]{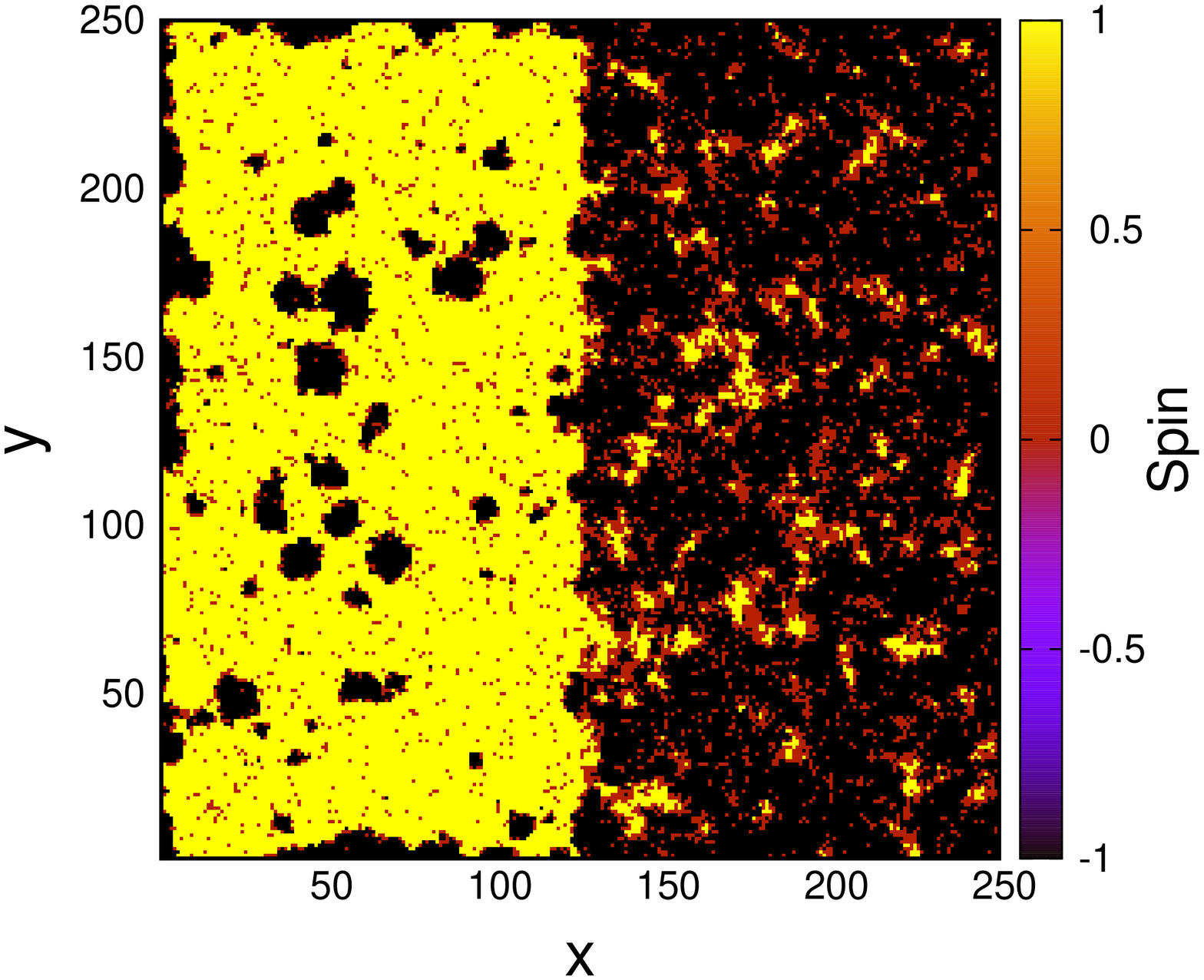}
		\subcaption{}
	\end{subfigure}
	\caption{Snapshots are taken for 3 different stepped anisotropic (where anisotropy act competitively with field) system at the time of reversal in presence of a 
		fixed $S_h$. $h_{sl}= -0.8$ and $h_{sr}= -0.4$; so the step difference of the applied field is $S_h= 0.4$. (a) $D_{sl}= 0.4$, $D_{sr}= 1.0$,  
		$S_D= 0.6$, $\tau= 123$ MCSS, (b) $D_{sl}= 0.4$, $D_{sr}= 1.32$, $S_D= 0.92$ (for which $C_F \sim 0.5$ at the $\tau$), 
		$\tau= 86$ MCSS, (c) $D_{sl}= 0.4$, $D_{sr}= 1.6$,  $S_D= 1.2$, $\tau= 51$ 
		MCSS. Temperature is set to $\textbf{T= 0.8}$.}
	\label{mcomp_snapsh}
\end{figure}

\begin{figure}[h!]
	\centering
		\includegraphics[angle=-90,width=0.5\textwidth]{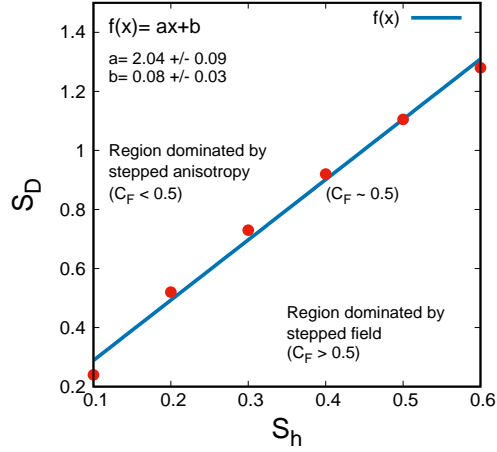}
	\caption{Relation of competitive stepped anisotropy having step difference $S_D$ for which 
		$C_F \sim 0.5$ with step difference $S_h$ of the applied stepped field.}
	\label{mcomp_sh}
\end{figure}

Since the reversal time of a system is largely dependent on the applied field and 
also on the anisotropy of the system, an interesting study (to incorporate the competitive behaviour of graded field and graded anisotropy) can be worked out on the behaviour 
of the system possessing graded anisotropy under the influence of a graded field. A gradient 
of field $G_h= 0.0016$ towards right edge is set up  by applying $h_l= -0.8$ and $h_r= -0.4$. 
So higher field near the left edge will try to flip the spins. In contrast if we take a graded 
anisotropic system having anisotropy $D_l=0.4$ and that at right edge is set to some higher value 
($D_r > D_l$), then the stronger anisotropy near right edge will also try to flip the spins. 
As a result, there would be a competitive scenario regarding reversal influenced by 
both field and anisotropy simultaneously. To explore this competitive behaviour quantitatively, 
a competition factor \cite{ranajay} can be defined as,
\begin{equation}
C_F = \frac{\big( \sum_{i=1}^{L^2} R_i \ast \sigma_i \big) + L^2}{2L^2}
\end{equation} 
where $R_i$ is the spin at i-th lattice site of a \textit{reference lattice} \cite{ranajay} whose left 
half sites are filled with spin '-1' only and the right half by spin '+1'. $\sigma_i$ is the spin at 
i-th lattice site of the main lattice we are studying. Now obviously at $\tau$, 
if the left side of the main lattice is dominated by spin '-1' and the right by spin '+1' 
(field dominated reversal) then $\big( \sum_{i=1}^{L^2} R_i \ast \sigma_i \big) \rightarrow 
L^2$, so $C_F \rightarrow 1.0$. On the other hand if the left side of the lattice is 
dominated by spin '+1' and the right by spin '-1' (anisotropy dominated reversal) then 
$\big( \sum_{i=1}^{L^2} R_i \ast \sigma_i \big) \rightarrow -L^2$, so $C_F \rightarrow 0$. 
If the small clusters of spin '1' and '-1' are equally and randomly distributed throughout the lattice then 
$\big( \sum_{i=1}^{L^2} R_i \ast \sigma_i \big) \rightarrow 0$, so $C_F \rightarrow 0.5$. 
Keeping the $G_h$ fixed, some snapshots are taken (fig-\ref{mcomp_snapgh}) for a single 
sample \textbf{at the time of reversal} for 3 different $G_D$ (acting competitively with 
applied field). In fig-\ref{mcomp_snapgh}a reversal is clearly dominated by graded field where 
$C_F > 0.5$ whereas in fig-\ref{mcomp_snapgh}c reversal is dominated by the anisotropy 
where $C_F < 0.5$. But in the fig-\ref{mcomp_snapgh}b field and anisotropy equally participate in 
reversal possessing  $C_F \simeq 0.5$ which is set as the benchmark of marginal competition. Now the value of $C_F$ is averaged over 100 samples and its 
variation with time (fig-\ref{cftime}) is also checked for the 3 cases. The arrow denotes the 
reversal time for each case ($C_F$ need not to be maximum at $\tau$). $C_F$ is almost constant 
($\simeq 0.5$) with time for equally competing field and anisotropy. By determining $C_F$ at $\tau$, the value of equally 
competing $G_D$ for several $G_h$ are found out by simple 
trial-and-error method and studied the variation  (fig-\ref{mcomp_gh}) of them. It follows a straight line 
$G_D \sim 1.88\;G_h $. We can call this line as the \textbf{'line of marginal competition'} 
 below which field plays dominating role in reversal and above which anisotropy plays dominating role. Same study (fig-\ref{mcomp_snapsh}\;\ref{mcomp_sh}) have been explored over a stepped anisotropic system in presence of a 
 stepped field where equally competing step differences $S_D$ and $S_h$ 
 are plotted. It also follows a straight line $S_D \sim 2.04\;S_h$.

\begin{figure}[h!]
	\begin{subfigure}{0.5\textwidth}
		\includegraphics[angle=-90,width=\textwidth]{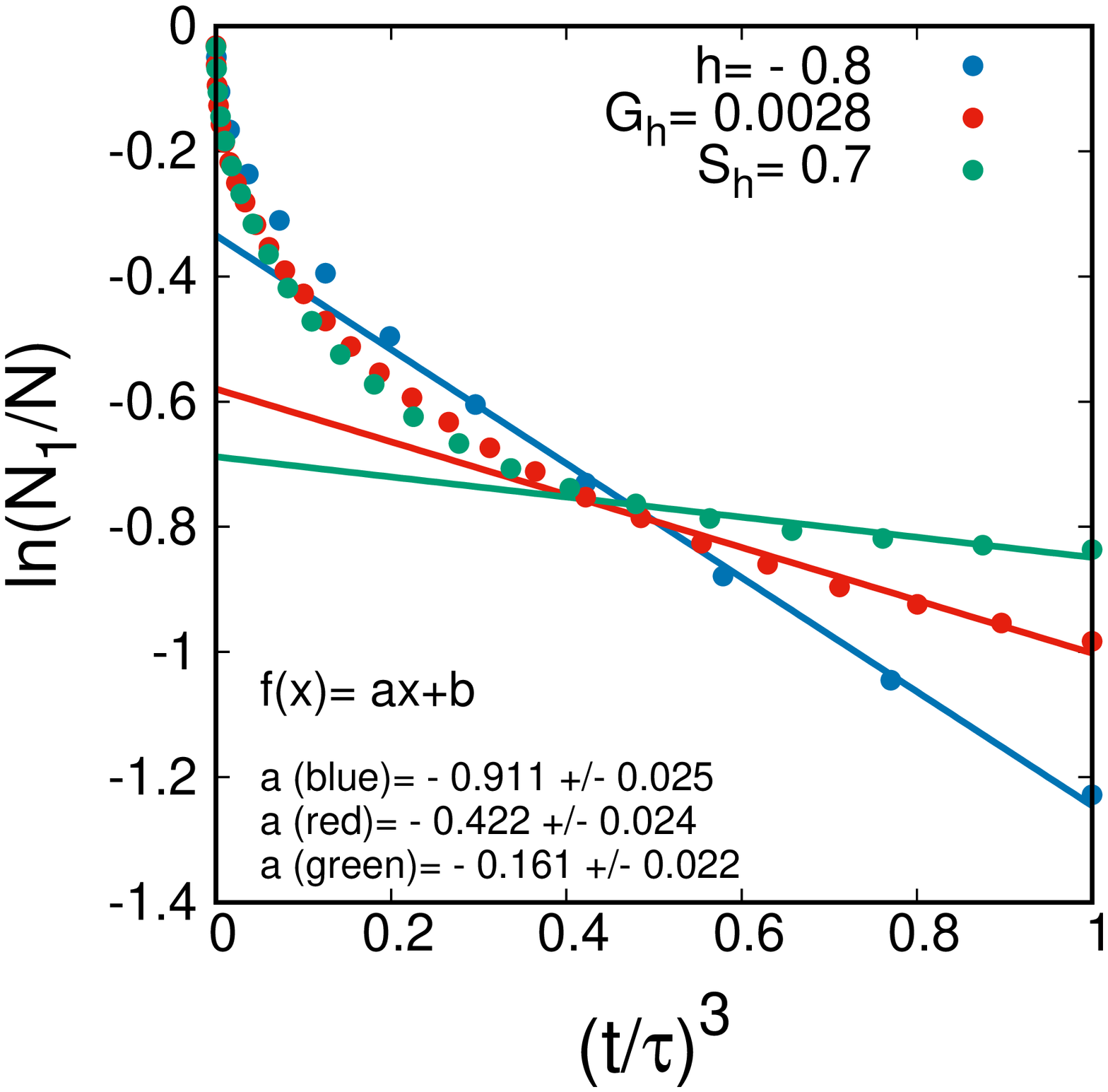}
		\subcaption{}
	\end{subfigure}	
	\begin{subfigure}{0.5\textwidth}
		\includegraphics[angle=-90,width=\textwidth]{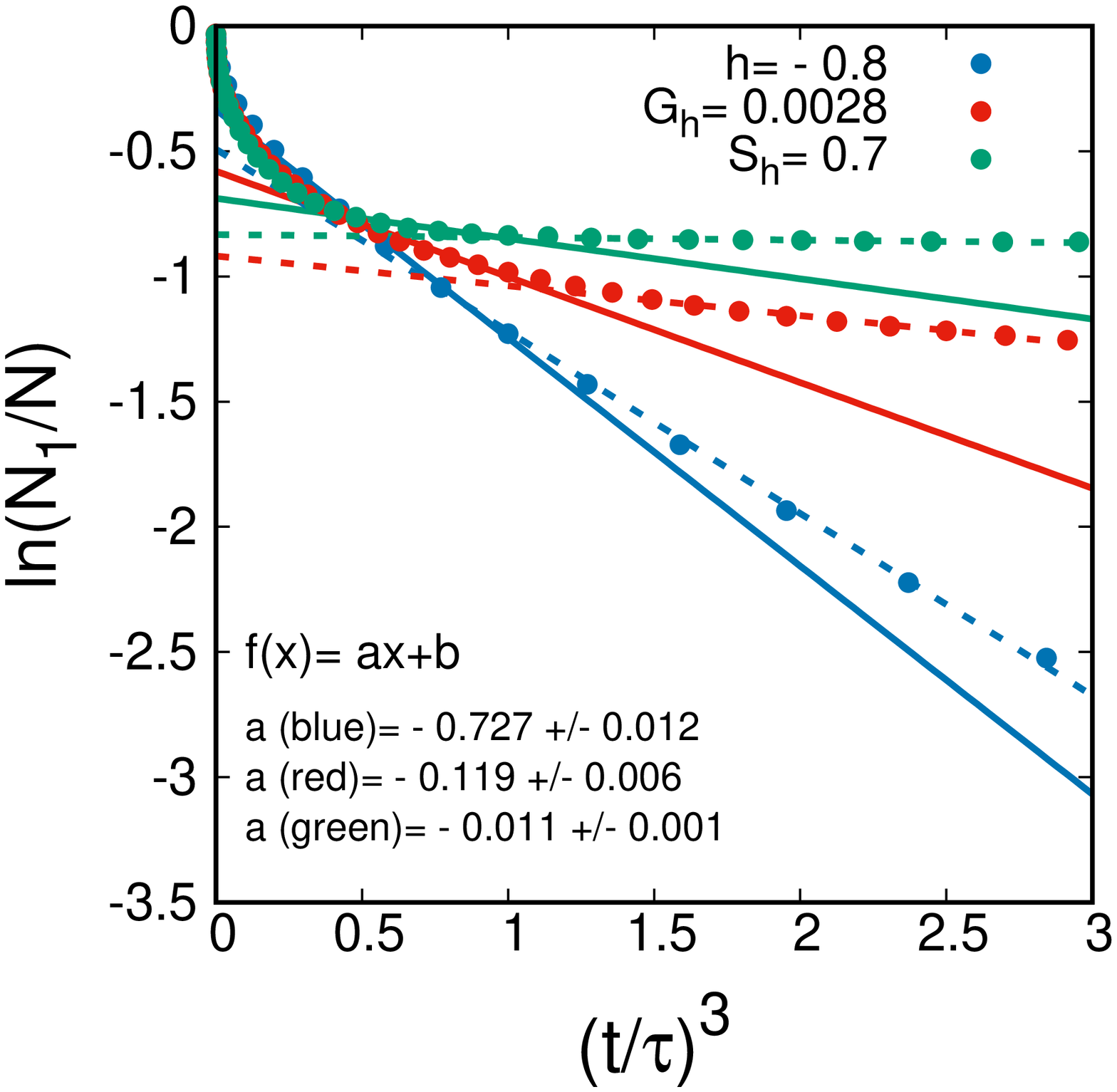}
		\subcaption{}
	\end{subfigure}
	\caption{Variation of logarithmic metastable volume fraction $\frac{N_1}{N}$ ($N_1$ is the number of spin +1, N is the total number of spin) with third power of time is plotted (a) upto reversal time $t=\tau$ where data are fitted near $\tau$. (b) upto $t= 3^{1/3} \tau$ where the solid lines are the same lines as in '(a)'. Now the data are fitted in the later times which are represented by the dotted lines. Anisotropy of the system is  
		$\textbf{D=1.6}$. Temperature is set to 
		$\textbf{T= 0.8}$.}
	\label{avrami}
\end{figure}

What will be the temporal behaviours of metastable volume fraction? Avrami’s law \cite{avrami} 
regarding the decay of metastable volume fraction has been investigated here for the anisotropic system in presence of 3 
different kinds of field. According to 
this law, for a d-dimensional system (closer to the critical temperature $T=0.8\;T_c$ here) 
metastable volume fraction (relative abundance of $\sigma_i^z$ = +1) decays exponentially 
with time $t^{d+1}$. So here (fig-\ref{avrami}a) the logarithm of the metastable volume fraction ($\frac{N_1}{N}$, $N_1$ 
is the number of spin '+1' and N is the total number of spin in the system) has been studied as function of 
the third power of dimensionless time 
$(\frac{t}{\tau})^3$ at temperature $T= 0.8$ \cite{butera}.

\begin{figure}[h!]
	\begin{subfigure}{0.5\textwidth}
		\includegraphics[angle=-90,width=\textwidth]{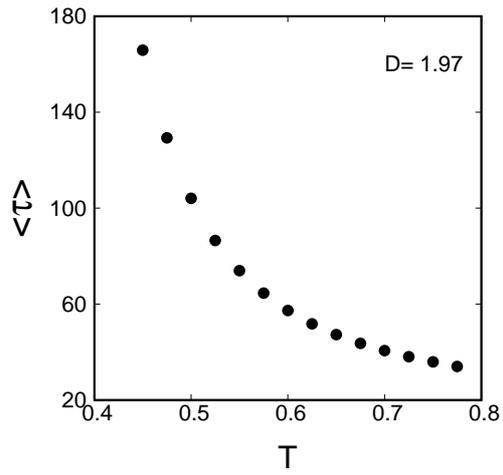}
		\subcaption{}
	\end{subfigure}	
	\begin{subfigure}{0.5\textwidth}
		\includegraphics[angle=-90,width=\textwidth]{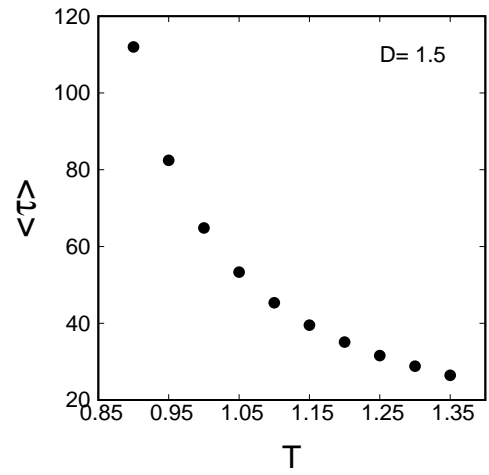}
		\subcaption{}
	\end{subfigure}
	\caption{Variation of reversal time across the (a) first order transition and (b)
	across the second oder transtion of the phase boundary. Here, applied field $h= -0.2$.}
	\label{nicos}
\end{figure}

How does the thermal variations of the reversal time look like if one crosses the
phase boundary\cite{nicos1,nicos2} of the BC model ? Can one observe any significant
change in the temperature dependence of the reversal time in the vicinity of the
phase transition ? To get the answers of these questions, we have studied the temperature dependence of the mean reversal time across the phase boundary 
(in the $T-D$ plane) of the
phase transitions studied\cite{nicos1,nicos2} recently. The results shown in 
Fig.\ref{nicos}(a) (where $D=1.97$, Fig.1 of Ref\cite{nicos2}) that if one crosses the first order transition
boundary, no significant changes in the thermal variation of reversal time is 
observed. The similar observation was made in the case of second order transition.
Fig.\ref{nicos}(b) shows that the thermal variation of mean reversal time is
insensitive to the second order transition boundary (here $D=1.5$, see Fig.1 of Ref\cite{nicos2}).

\vskip 2 cm
\noindent {\large\bf IV. Summary} 
\vskip 0.2 cm
In this article, the behaviours of metastable lifetimes and the switching of magnetisation in the  anisotropic Blume-Capel ferromagnet has been studied 
by extensive Monte Carlo simulation using Metropolis single spin flip algorithm.
The switching mechanism is governed generally by the application of 
uniform external magnetic field. How does the switching behave in the spatially
modulated external magnetic field and the anisotropy of the system, is the
main objective of this study. 

In the first part, the two different kinds of spatial modulation (namely graded and
stepped like) of external magnetic field are considered with uniform anisotropy of
the system. The reversal time (or switching time) was found to increase in the case
of graded field as compared to that for uniform field. In the case of graded field 
a major portion of the lattice experiences weaker magnetic field which plays the
role of such delayed reversal. Here, an interesting phenomenon was observed in support
of classical nucleation theory (or Becker-Doring theory). Considering two small parts
of the whole lattice where the fields are relatively weaker and relatively stronger.
In the microlattice where the field is relatively weaker, the growth of single nucleating cluster was observed. On the other hand, in the microlattice where the
applied field is relatively stronger, the coalescence of multiple droplets was 
observed. In the case of graded field the mean reversal time was found to follow
the $<\tau> \sim {\rm e}^{bG_{h}}$. A crossover in the value of the rate ($b$) 
of this exponential growth was observed from the lower value of the gradient to higher value of the gradient of the external magnetic field. In the case of stepped 
magnetic field, a linear growth of mean reversal time was observed. The dependence of
the mean reversal time on the system size was found to obey a scaling relation
$<\tau> \sim L^{-\beta}f(G_hL^{\alpha})$ with $\alpha \simeq 1.06$ and $\beta \simeq 0$.

The nonuniform spatial variation of the mean densities of spin projections was
observed in the case of graded magnetic field. The density of +1 was found to
decrease from left to right in contrary to that of -1. The density of 0 gets peaked
nearly in the middle. 

The mean numbers of spin flip also shows a nonuniform spatial variation in the
case of graded magnetic field. Interestingly, for higher gradient this shows a very
sharp peak in the middle. In the case of stepped field, this shows a finite discontinuity in the middle, clearly reflecting the stepped discontinuity of the
applied magnetic field.

The mean position and the roughness of the domain wall was studied as function of
the gradient  of the applied magnetic field. In the case of
graded field, the mean position of the domain wall was observed to shift towards right
in a hyperbolic tangent fashion with the value of the gradient and the roughness
of the interface of the domains was found to decay in a stretched exponential manner.

The role of spatial modulation of the anisotropy of the system was also studied by
using graded and stepped anisotropy where the external magnetic field was uniform
over the space. In the case of graded anisotropy, the mean reversal time $<\tau>$
was found to decrease exponentially ($<\tau> \sim {e}^{-bx}$) with the gradient ($G_D$) of the anisotropy. 
However, in the case of step like variation of the anisotropy, it was found to
decay in a stretched exponential manner ($<\tau> \sim {e}^{-b'x^{0.5}}$). 

Mean numbers of spin flips were found to increase from left to right side of the
lattice in the case of graded anisotropy and it shows finite discontinuity in the middle for the case of stepped anisotropy.

The mean position and the roughness of the domain wall was studied as function of
the gradient  of the anisotropy. In the case of
graded anisotropy, the mean position of the domain wall was observed to shift towards right
in a hyperbolic tangent fashion (like the case of graded field) with the value of the gradient and the roughness
of the interface of the domains was found to decay exponentially (unlike the case of
graded field).

Having the knowledge of the reversal of magnetisation by graded field and
graded anisotropy, the joint effects of both were investigated. Here, a competetive
scenario was observed and the line of marginal competetion was drawn in the plane
of gradients of field and anisotropy. The linear boundary was found to separate the
regions of field dominated reversal and anisotropy dominated reversal.

The joint effects of stepped field and anisotropy were investigated. Here also, like the case of graded variations, a competetive
scenario was observed and the line of marginal competetion was drawn in the plane
of step discontinuities of field and anisotropy. The linear boundary was found to separate the regions of field dominated reversal and anisotropy dominated reversal.

The decay of metastable volume fraction was studied in the case of uniform,
graded and stepped field to check the Avrami kind of variation
\cite{ramos}. In all cases the
the logarithmic volume fractions were found to decay as $-(t/\tau)^3$ supporting
Avrami behaviours. However, the decay becomes relatively slower (with smaller decay rate) after the reversal.

Finally, we have studied the temperature dependence of mean reversal time across
the phase boundary of first order and second order ferro-para phase transitions\cite{nicos2}. We have
not observed any significant change (near the transition) in the behaviour of the monotonic 
decrease of the reversal time as one moves from ferro to para region.

Let us make a few remarks on the possible potential application and experimental objective of the present study. The
reversal time is important in switching phenomena in magnetic devices. This reversal time
generally depends on the material which cannot be changed from external agency. 
However, the
experimentalist can think of an alternative method of tuning the reversal time by means of varying the
gradient of the external fields.
\vskip 2 cm
\noindent {\large\bf V. Acknowledgements}
 
MN would like to acknowledge Swami Vivekananda Scholarship (SVMCMS) for financial support. MA acknowledges FRPDF grant of Presidency University for financial support.

\vskip 0.2 cm


\vskip 2 cm
\newpage

\newpage

\newpage


\newpage

\newpage

\newpage

\newpage

\newpage

\clearpage
\newpage

\newpage		

\newpage


\newpage

\newpage

\newpage

\newpage

\newpage

\newpage
\begin{table}[ht]
	\caption{Goodness of fit $\chi^2$ statistics\cite{young} of Fig.\ref{if_gh} and Fig.\ref{if_gd}}
	\centering
	\begin{tabular}{c c c c}
		\hline\hline
		Figure & value of $\chi^2$ & Degrees of freedom & Probability(Q-value)\\
		\hline
		Fig.\ref{if_gh}(a) & 3.9514 & 6 & 0.6832\\
		Fig.\ref{if_gh}(b) & 0.3086 & 6 & 0.9994\\
		Fig.\ref{if_gd}(a) & 5.5372 & 5 & 0.3539\\
		Fig.\ref{if_gd}(b) & 5.6463 & 5 & 0.3422\\
		\hline
    \end{tabular}
\end{table}

\end{document}